\title{Power and Energy Models for DSR1 Series}
\documentclass[conference]{IEEEtran}
\IEEEoverridecommandlockouts
\usepackage{cite}
\usepackage{pgfplots}

\usepackage{amsmath,amssymb,amsfonts}
\usepackage{algorithmic}
\usepackage{graphicx}
\usepackage{adjustbox}
\usepackage{subcaption}
\usepackage{textcomp}
\usepackage{xcolor}
\usepackage{placeins}
\usepackage{amssymb}
\usepackage{siunitx}
\usepackage{booktabs} 
\usepackage{siunitx}  
\usepackage{float}
\usepackage{tikz}
\usepackage{tabularx}
\usepackage{listings}
\usepackage{url}
\usepackage[colorlinks=true, linkcolor=blue, citecolor=blue, urlcolor=blue,]{hyperref}
\usepackage{multirow} 
\definecolor{skinlight}{RGB}{253, 236, 222}
\definecolor{skinborder}{RGB}{155, 100, 75}
\usepackage{tcolorbox}
\newtcolorbox{keytakeaway}{
    colback=skinlight,        % Light blue background
    colframe=skinborder,      % Dark blue frame
    fonttitle=\bfseries,
    rounded corners,               % Square corners
    boxrule=1pt,                 % Border thickness
    left=2mm,                    % Left padding
    right=2mm,                   % Right padding
    top=2mm,                     % Top padding
    bottom=2mm,                  % Bottom padding
    boxsep=5pt,                  % Inner spacing
    before skip=12pt,            % Space before box
    after skip=12pt              % Space after box
}

\definecolor{DSR1-Qwen-1.5B}{HTML}{2E86AB}
\definecolor{DSR1-Llama-8B}{HTML}{A23B72}
\definecolor{DSR1-Qwen-14B}{HTML}{F18F01}
\definecolor{DS-Qwen-1.5B}{HTML}{2E86AB}
\definecolor{DS-Llama-8B}{HTML}{A23B72}
\definecolor{DS-Qwen-14B}{HTML}{F18F01}
\definecolor{L1-Max}{HTML}{7FB069}
\def\BibTeX{{\rm B\kern-.05em{\sc i\kern-.025em b}\kern-.08em
    T\kern-.1667em\lower.7ex\hbox{E}\kern-.125emX}}
\begin{document}

\newcommand{\symonetwoeightT}{$\Diamond$}      % Bold cross
\newcommand{\symonetwoeightTNC}{$\square$}            % Diamond
\newcommand{\symtwofivesixT}{$\triangle$}             % Triangle
\newcommand{\symtwofivesixTNC}{$\triangledown$}       % Downward triangle
\newcommand{\symBase}{$\circ$}                 % Circle
\newcommand{\symNR}{$\star$}                   % Star
\newcommand{\symDirect}{$+$}   

\newcommand{\jenny}[1]{{\color{blue} {Jenny: #1}}}
\newcommand{\todo}[1]{{\color{red} {TODO: #1}}}

\newcommand{\sys}[0]{\textcolor{black}{EdgeReasoning{}}}

%page number
% \pagestyle{plain} 
%When Thinking Out Loud Hurts: Measuring Chain-of-Thought Latency on Edge GPUs
% \title{Characterizing Latency of Reasoning LLMs On Edge GPUs\\
% }
% \title{Characterizing Optimal LLMs Deployment for Reasoning Tasks on Edge GPUs\\
% } 
\title{\sys: Characterizing Reasoning LLM Deployment on Edge GPUs\\
} 
\author{\IEEEauthorblockN{Benjamin Kubwimana}
\IEEEauthorblockA{\textit{NVIDIA} \\
% Santa Clara, USA \\
bkubwimana@nvidia.com}
\and
\IEEEauthorblockN{Qijing Huang}
\IEEEauthorblockA{\textit{NVIDIA} \\
% Santa Clara, USA \\
jennyhuang@nvidia.com}
}

\maketitle

\begin{abstract}

% LLMs have achieved unprecedented success in reasoning tasks. However, deploying them on edge devices is challenging due to limited capacity constraints and tight real-time requirements.
% When setting up inference tasks for edge deployment of reasoning model, there are several critical tradeoffs to consider: 
% choosing among reasoning vs nonreasoning models, different model sizes, various token length budgets, and test time scaling methods. 
% It is unclear how to choose the optimal combination to achieve optimal accuracy under various latency requirements. In this paper, 
% we first characterize the latency and accuracy tradeoff among non-reasoning and reasoning models. 
% Then we analyze various prompt based optimization to reduce the inference latency of reasoning models as well test time scaling methods to maximize accuracy under inference latency constraints. 

Edge intelligence paradigm is increasingly demanded by the emerging autonomous systems, such as robotics. %, such as self-driving and humanoid robots, , drones, 
Beyond ensuring privacy-preserving operation and resilience in connectivity-limited environments, edge deployment offers significant energy and cost advantages over cloud-based solutions. However, deploying large language models (LLMs) for reasoning tasks on edge GPUs faces critical challenges from strict latency constraints and limited computational resources.

To navigate these constraints, developers must balance multiple design factors—choosing reasoning versus non-reasoning architectures, selecting appropriate model sizes, allocating token budgets, and applying test-time scaling strategies—to meet target latency and optimize accuracy. Yet guidance on optimal combinations of these variables remains scarce. 
% choosing reasoning versus non-reasoning architectures,
% forcing practitioners to rely on ad-hoc approaches that may significantly underperform.

In this work, we present \sys{}, a comprehensive study characterizing the deployment of reasoning LLMs on edge GPUs. We systematically quantify latency-accuracy tradeoffs across various LLM architectures and model sizes. We systematically evaluate prompt-based and model-tuning-based techniques for reducing reasoning token length while maintaining performance quality.  We further profile test-time scaling methods with varying degrees of parallelism to maximize accuracy under strict latency budgets. Through these analyses, \sys{} maps the Pareto frontier of achievable accuracy-latency configurations,  offering systematic guidance for optimal edge deployment of reasoning LLMs.

\end{abstract}

\section{Introduction}

The rapid advancement of autonomous systems—from robotics and drones to self-driving vehicles—has created an unprecedented demand for intelligent decision-making and reasoning capabilities at the edge~\cite{zhou2019edge}. Consider personal assistive humanoid robots: when a user requests ``Can you help me prepare dinner within 5 minutes?'', the robot must perform real-time planning and execution under strict latency constraints. Such scenarios reveal a critical tension - tasks with generous latency budgets (e.g., ``Plan my weekly schedule'') benefit from larger models with longer reasoning chains for optimal planning, while latency-sensitive tasks (e.g., ``Avoid that obstacle now!'') demand smaller models that sacrifice optimality for speed. 

\begin{figure}[t]
    \centering
    \includegraphics[width=0.95\linewidth]{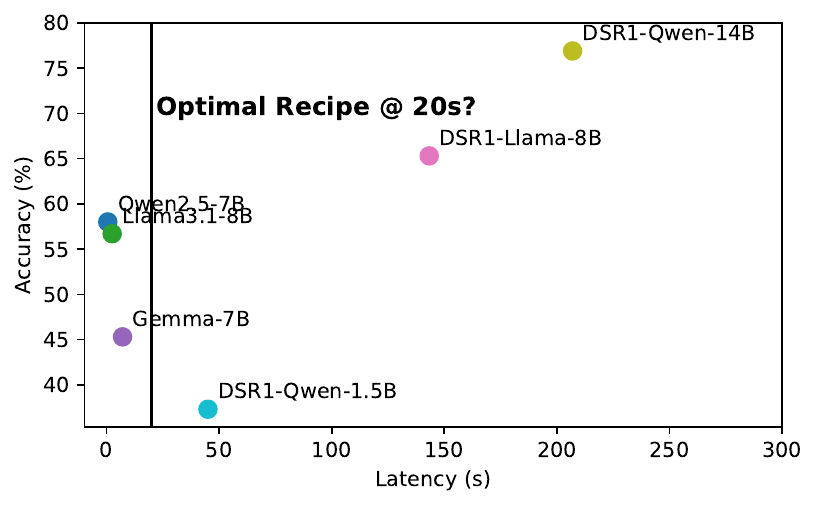}
    \caption{Discrete accuracy-latency tradeoffs fail to capture continuous operational requirements of real-world systems like assistive robots.}
    \label{fig:movtivation}
\end{figure}

This operational reality presents fundamental challenges for edge deployment of reasoning models. First, the autoregressive nature of LLMs creates highly variable token generation times, making latency hard to control---potentially resulting in missed deadlines or no responses. 
Second, deploying reasoning-capable models incurs substantial latency dominated by decoding processes, particularly problematic for real-time systems.
Third, the discrete accuracy-latency tradeoffs shown in Fig.~\ref{fig:movtivation} fail to capture the continuous spectrum of real-world requirements. 
These challenges necessitate: (1) precise token length control to meet latency constraints, (2) hardware-aware functions mapping latency budgets to maximum decodable tokens, and (3) continuous optimization across the latency-accuracy frontier.

While cloud-based large language models (LLMs)~\cite{guo2025deepseek, jaech2024openai, team2023gemini} have demonstrated remarkable reasoning abilities, the edge intelligence paradigm offers compelling advantages that extend far beyond privacy preservation and connectivity resilience. Most notably, edge deployment presents transformative cost efficiencies that alter the economics of AI-powered autonomous systems. Recent developments in lightweight reasoning models~\cite{luo2025deepscaler, deepcoder2025} have achieved comparable accuracy to larger commercial models at two orders of magnitude lower cost, as demonstrated in Section~\ref{sec:cost}.

% However, realizing these advantages while maintaining responsiveness presents significant technical challenges. Edge GPUs impose strict computational constraints that complicate LLM deployment. The complex interplay between model architecture, token budget allocation, and test-time scaling creates a vast design space where suboptimal decisions undermine both performance and efficiency. As illustrated in Fig.~\ref{fig:movtivation}, current approaches to edge LLM deployment exist as isolated solutions, leaving practitioners without systematic guidance for navigating critical tradeoffs.

However, realizing these cost advantages while maintaining reasoning quality presents significant technical challenges. 
Edge deployment imposes latency constraints and computational limitations that complicate the inference strategies of reasoning LLMs. The complex interplay between model architecture choices, token budget allocation, and test-time scaling strategies creates a vast design space where suboptimal decisions can undermine both performance and efficiency.
As illustrated in Fig.~\ref{fig:movtivation}, current approaches to edge LLM deployment exist as isolated solutions, leaving practitioners without systematic guidance for navigating the critical tradeoffs between reasoning depth, inference speed, and accuracy. This gap is particularly problematic for autonomous systems where both real-time responsiveness and reliable decision-making are essential requirements.

In this work, we address these challenges through \sys, a comprehensive study for characterizing LLM reasoning deployment on edge GPUs. Our contributions include: 
\begin{enumerate}
    \item Empirical characterization of latency, power, and energy tradeoffs across LLM architectures on edge hardware.
    \item Performance models that analytically maps token counts to latency and energy performance for edge GPUs. 
    \item Systematic exploration of prompt-based optimization techniques to reduce reasoning token overhead.
    % \item Novel methods for continuous mapping of latency budgets to optimal model configurations.
    \item Evaluation of test-time scaling methods to maximize accuracy under dynamic latency constraints.
\end{enumerate}
Through this study, we demonstrate how edge reasoning can achieve cost efficiencies that make autonomous AI systems economically sustainable while providing deterministic latency control essential for real-time applications. Crucially, \sys{} study enables autonomous systems to select optimal accuracy configurations within task-specific latency requirements, maximizing performance across diverse operational scenarios.
% \clearpage
\section{Background} %Jenny
\label{sec:background}
\subsection{Reasoning LLMs} % model size 
\label{sec:reasoning-llms}
Recent advances in large language models have enabled multistep logical reasoning and complex problem solving capabilities. Reasoning LLMs (e.g., OpenAI o1~\cite{jaech2024openai}, DeepSeek-R1~\cite{guo2025deepseek}) generate intermediate ``chains of thought'' (CoT)~\cite{wei2022chain} that decompose complex problems into sequential inference steps before producing final answers. These models achieve superior accuracy on challenging tasks, including mathematics and coding, compared to traditional direct-generation counterparts. However, reasoning LLMs generate significantly longer output sequences than non-reasoning models, creating substantial computational overhead for edge deployment.

% They also demonstrate emergent behaviors such as self-correction and backtracking during inference, allowing dynamic adjustment of reasoning paths when initial approaches prove inadequate.
To address edge deployment constraints, \emph{lightweight language models} have been developed~\cite{team2024gemma,yang2025qwen3,abdin2024phi}.  Complementing these, knowledge distillation from large reasoning LLMs followed by task-specific fine-tuning has yielded compact reasoning models that retain high accuracy. For example, DeepSeek-R1~\cite{guo2025deepseek} is available in 1.5B, 7B, 8B, and 14B parameter variants optimized for edge devices. Additionally, DeepScaleR/DeepCodeR~\cite{luo2025deepscaler, deepcoder2025}, fine-tuned with reinforcement learning, attains parity with large models such as OpenAI’s o1 on mathematical and coding tasks—demonstrating that sub 15B models can deliver near state-of-the-art reasoning performance within edge-scale compute and memory budgets.

\subsection{Test-Time Scaling} % parallel scaling factor 
\label{sec:test-time-scaling}
Recent work has shifted focus from training-time scaling to test-time scaling, allowing LLMs to ``think with more tokens''~\cite{jaech2024openai, snell2024scaling}. Test-time scaling laws demonstrate predictable accuracy gains from increased inference computation through generating more or longer reasoning chains. Test-time scaling can be achieved through two main approaches: \emph{sequential scaling}~\cite{muennighoff2025s1}, which extends the length of individual reasoning chains, and \emph{parallel scaling}~\cite{brown2024large},  where multiple reasoning paths are generated simultaneously across processing units and aggregated via voting or consensus mechanisms.  %codemonkey 
While both approaches multiply computational requirements, parallel scaling avoids linear latency increases through parallelization, making it particularly attractive when hardware resources are underutilized. 
More sophisticated inference strategies integrate both sequential and parallel scaling~\cite{gao2024interpretable, huggingface2024scaling, hooper2025ets, wu2024inference}.
% There are more complex inference strategies that combine both scaling using tree search such as ~\cite{gao2024interpretable, huggingface2024scaling, hooper2025ets, wu2024inference}.

% MCTS~\cite{gao2024interpretable}, DVTS~\cite{huggingface2024scaling}, ETS~\cite{hooper2025ets}, and REBASE~\cite{wu2024inference}. %MCTS and REBASE 

% \subsection{Efficient LLMs}
% \label{sec:efficient-models}
% Deploying LLMs under tight latency and memory budgets requires techniques that reduce model footprint while preserving inference quality. Core optimization methods include quantization (reducing weight precision to INT8/FP8 for lower memory usage and hardware acceleration), knowledge distillation (training smaller student models to mimic larger teachers)~\cite{luo2025deepscaler, guo2025deepseek}, and pruning with early exit (removing redundant parameters and terminating generation when confidence is high).
% Beyond these standard approaches, deployment-specific strategies can further optimize performance. Output length limitation directly controls latency since generation length dominates inference time—producing 50 tokens is roughly twice as fast as 100 tokens. Task complexity management involves restricting edge models to simple queries while offloading complex multi-step reasoning to cloud services, effectively capping worst-case latency by avoiding computationally expensive deep reasoning scenarios on resource-constrained devices. 

\subsection{Reasoning Token Optimization} % length control

Besides employing lightweight models and parallel test-time scaling techniques, 
optimizing reasoning length offers another approach to deploy reasoning models under latency constraints while preserving accuracy.~\cite{sui2503stop} 
% pruning redundant output tokens and reasoning steps 
This can be achieved through
\emph{prompt-based methods}, which instruct models to use fixed token budgets \cite{han2024token} or disable CoT reasoning \cite{ma2025reasoning}, trading off reasoning depth for reduced latency. However, these approaches are limited as not all models are trained with token budget awareness.

Alternatively, \emph{fine-tuning techniques} like length-difference positional encoding \cite{butcher2025precise} or explicit output length control (e.g., L1 \cite{aggarwal2025l1}) achieve precise sequence length control. 
While effective at reducing output length for reasoning, these methods lack system-level integration to show practical latency-accuracy improvements in real deployments. See also latency-aware test-time scaling
\cite{wang2025faster}.
 
% Besides employing lightweight models and efficient TTS techniques, pruning redundant output tokens and the reasoning process in token generation is another way to enable reasoning LLMs under latency constraints while preserving inference accuracy.
% This can be achieved through prompting to instruct the model to use a fixed number of tokens in the response \cite{han2024token, sui2503stop, muennighoff2025s1}, or to disable CoT reasoning \cite{ma2025reasoning}, trading reasoning depth for shorter latency.
% However, not all models are trained with token budget awareness. \cite{butcher2025precise} fine-tuned the model and incorporated length-difference positional encoding to achieve precise length control. L1 \cite{aggarwal2025l1} has managed to control the output length of reasoning models.
% While these works successfully limit sequence length control, they did not show how they can be integrated into real systems to achieve better latency and accuracy performance.
 
\begin{table}[t]
\centering
\caption{NVIDIA Jetson Orin Series Compute Specifications}
\label{table:orin}
\resizebox{1\columnwidth}{!} {

\begin{tabular}{|l|l|l|l|}
\hline
\textbf{CUDA Cores}  & \textbf{Tensor Cores} & \textbf{DLA} & \textbf{Memory} \\ \hline
2048 (5.3TFLOPs) & 64 (275TOPs) & 2 (52.5TOPS) & 64GB @ 204.8GB/s \\
% \texttt{Orin NX 16GB} & 1024 (32) & 1× NVDLA v2 & 16GB LPDDR5 \
% \texttt{Orin Nano 8GB} & 512 (16) & — & 8GB LPDDR5 \
\hline
\end{tabular}}
\end{table}  
\subsection{Edge GPUs}
\label{sec:edge-gpu}
Deploying a reasoning LLM on an edge device, such as NVIDIA Jetson AGX Orin\cite{nvidiaJetsonOrin}, imposes strict latency and memory constraints. Edge GPUs have limited compute throughput and memory bandwidth and capacity compared to server accelerators, making the lengthy decode phase of reasoning LLMs especially challenging.

NVIDIA’s Jetson AGX Orin, which we use for all studies in this paper, is a representative edge AI GPU platform that integrates advanced compute capabilities in a low-power package. As shown in Table~\ref{table:orin}, the Orin system-on-chip (SoC) features an NVIDIA Ampere-architecture GPU with 2048 CUDA cores, 64 Tensor Cores, and 2 NVDLAv2 Cores.  The Tensor Cores accelerate mixed-precision matrix operations, enabling high-throughput FP16 and INT8 computations for deep learning acceleration. The Jetson Orin’s GPU can deliver up to roughly 5.3 TFLOPs of FP32 compute or up to 275 Sparse INT8 TOPS for deep learning workloads. The memory hierarchy includes 4MB of GPU L2 cache and 3MB of aggregate GPU L1 cache (192KB $\times$ 16 SMs). The platform features 64GB of LPDDR5 memory and operates within a configurable power envelope of 15–60W, making it well-suited for embedded applications in robotics and autonomous driving. The GPU is complemented by a 12-core ARM Cortex-A78AE CPU for control-heavy processing tasks.

% Total prompt tokens: 3782
% Total completion tokens: 191842
% Total electricity cost: \$0.00475828
% Cost per 1M  tokens: \$0.0248

\begin{table}
\caption{Comparison of Lightweight Reasoning and Non-Reasoning Models for 150 MMLU-Redux Questions.}

\label{tab:non-reasoning}
\resizebox{1\columnwidth}{!} {
    \centering
    \begin{tabular}{cccccc}\toprule
         Model&  Acc. (\%)& Time (s)&  TPS&  Perf/W& Energy/Q (J)\\\midrule
         gemma-7B~\cite{team2024gemma}&  33.9 &  7.1&  7.2&  0.3& 210.3
\\
         llama3.1-8B~\cite{grattafiori2024llama}&  58.3&  2.5&  6.6&  0.3& 77.9
\\
         qwen2.5-7B~\cite{qwen2.5}&  60.8&  0.6&  7.2&  0.3& 26.4
\\
         DSR1-Qwen-1.5B~\cite{guo2025deepseek}&  38.3&  45.0&  9.3&  1.1& 403.6
\\
 % L1-Qwen-1.5B-Max~\cite{aggarwal2025l1}& 44.1& 19.3& 16.0& 1.1&281.2\\
         DSR1-LLama-8B~\cite{guo2025deepseek}&  61.7&  143.3&  7.8&  0.3& 4205.5
\\
         DSR1-Qwen-14B~\cite{guo2025deepseek}&  80.6&  207.0&  4.7&  0.2& 2599.2
\\ \bottomrule
    \end{tabular}
}
\end{table}

\section{Motivation}
\subsection{Comparison of Reasoning vs Non-Reasoning LLMs}

% Table~\ref{tab:non-reasoning} compares reasoning and non‐reasoning models across key metrics: MMLU-Redux accuracy, average decoding time, tokens per second (TPS), performance per watt, and energy per question.  We evaluate distilled reasoning models from the DeepSeek-R1~\cite{deepseek_r1_0528} family against popular lightweight non-reasoning alternatives (Gemma~\cite{team2024gemma}, Llama3~\cite{grattafiori2024llama}, Qwen2.5~\cite{qwen2.5}).

% On 150 MMLU-Redux~\cite{gema2024mmlu} questions, reasoning models deliver substantially higher accuracy—scaling positively with model size—and outperform non‐reasoning models of similar scale (7–8B parameters) by over 7\%. However, this accuracy boost incurs a steep cost: reasoning models run over 20× slower on average, driving energy and per‐token cost up by at least a factor of 20.  
 
Table~\ref{tab:non-reasoning} presents a comparison between reasoning and non-reasoning models across multiple performance metrics: MMLU-Redux accuracy~\cite{gema2024mmlu}, average decoding time, tokens per second (TPS), performance per watt, and total energy consumption per question.
We evaluate distilled reasoning models from the DeepSeek-R1~\cite{deepseek_r1_0528} family and three popular lightweight non-reasoning alternatives (Gemma~\cite{team2024gemma}, Llama3~\cite{grattafiori2024llama}, Qwen2.5~\cite{qwen2.5}).
Reasoning models demonstrate substantially higher accuracy than their non-reasoning counterparts on MMLU-Redux benchmarks\cite{gema2024mmlu}. Furthermore, accuracy scales positively with model size among the distilled reasoning models.
When comparing models of similar scale (7-8B parameters), reasoning models achieve more than 7\% higher accuracy than non-reasoning alternatives. However, this performance gain comes with significant computational overhead:\textbf{ \emph{reasoning models incur over 20× higher inference latency than non-reasoning models}}. Consequently, both energy consumption and cost per token increase by a similar factor of at least 20$\times$ compared to non-reasoning counterparts. 
% Therefore, strategies to optimize output token length become critical for practical deployment.
These efficiency gaps underscore the need for output token optimization strategies for practical deployment of reasoning models at the edge. 

\begin{table}[t]
\centering
\caption{Costs Comparison of Reasoning LLM Deployments}
\label{tab:model-comparison}
\resizebox{0.9\columnwidth}{!} {
\begin{tabular}{lccc}
\toprule
\textbf{Metric} & \textbf{OpenAI o1-preview} & 
\multicolumn{2}{c}{\textbf{DeepScaleR-1.5B}} \\
\midrule
\textbf{Parameter Size} & Unknown & \multicolumn{2}{c}{1.5B in fp16} \\
\midrule
\textbf{Accuracy (AIME2024)} & 40.0\% & \multicolumn{2}{c}{\textbf{43.1\%}} \\
\textbf{Accuracy (Math500)} & 81.4\% & \multicolumn{2}{c}{\textbf{87.8\%}} \\
\midrule
\textbf{Batch Size} & Unknown & 1 & 30 \\
\midrule
\textbf{Throughput (User TPS)} & 89.7~\cite{openai_o1_preview} & 44.0 & 21.2 \\
\midrule
\textbf{Price (Input \$/1M tokens)} & \$15~\cite{openai_api_pricing} & \$0.302 & \textbf{\$0.027} \\
\textbf{Price (Output \$/1M tokens)} & \$60 & \$0.302 & \textbf{\$0.027} \\
\bottomrule
\end{tabular}}
\end{table}
% \todo{Add SOL cost analysis for edge}

\subsection{Comparison of Edge vs. Cloud Deployment}
\label{sec:cost}
Edge deployment provides substantial energy and cost savings compared to cloud solutions, while also ensuring data privacy and operational resilience in connectivity-constrained environments.
Table~\ref{tab:model-comparison} demonstrates the significant cost efficiency of deploying DeepScaleR-1.5B on edge devices like the NVIDIA Jetson AGX Orin compared to cloud-based APIs. While OpenAI’s reasoning models charge more than \$4 per million output tokens  (\$4.4 on o4-mini and \$60 on o1-preview), DeepScaleR achieves \$0.302 per million tokens by running entirely on-device. DeepScaleR-1.5B further excels in accuracy, outperforming the commercial cloud model o1-preview on the AIME2024 and MATH500 benchmarks due to its RL fine-tuning for math and reasoning. This shows that \textbf{\textit{edge deployment of reasoning models can deliver competitive accuracy at radically lower costs. }}

Cost calculations derived from profiling the AIME2024 benchmark on the Orin platform reveal: In single-batch processing (FP32), the system handled 195,624 tokens in 4,358~seconds using 0.0317~kWh. At electricity rates of \$0.15/kWh and hardware amortized at \$0.045/hour, this yields \$0.302 per million tokens (\$0.024 energy + \$0.278 hardware). Notably, batch processing (size~30) completed the same workload in 398~seconds using only 0.003~kWh, reducing costs to \$0.027 per million tokens (\$0.0023 energy + \$0.025 hardware). 
These results demonstrate that \textbf{\textit{edge deployment costs also benefit from batching and increased queries per second (QPS)}}. 
\section{Edge GPU Performance Characterization and Modeling}
In this section, we characterize the latency, power, and energy consumption of lightweight reasoning models across different sizes (1.5B, 8B, 14B) with respect to various prefill and decode lengths when deployed on the Jetson Orin GPU using vLLM \cite{kwon2023efficient} as the inference engine.

\subsection{Characterization of Latency}
\label{sec:latency}

We first analyze end-to-end inference latency, decomposing it into prefill (initial prompt processing) and decode (token generation) components to reveal distinct computational behaviors. Beyond empirical measurements, we develop and validate accurate analytical performance models fitted to real-world Jetson Orin measurements, enabling rapid evaluation and navigation of latency-accuracy trade-offs during inference strategy selection.

\begin{figure}[t]
    \centering
    \includegraphics[width=0.8\linewidth]{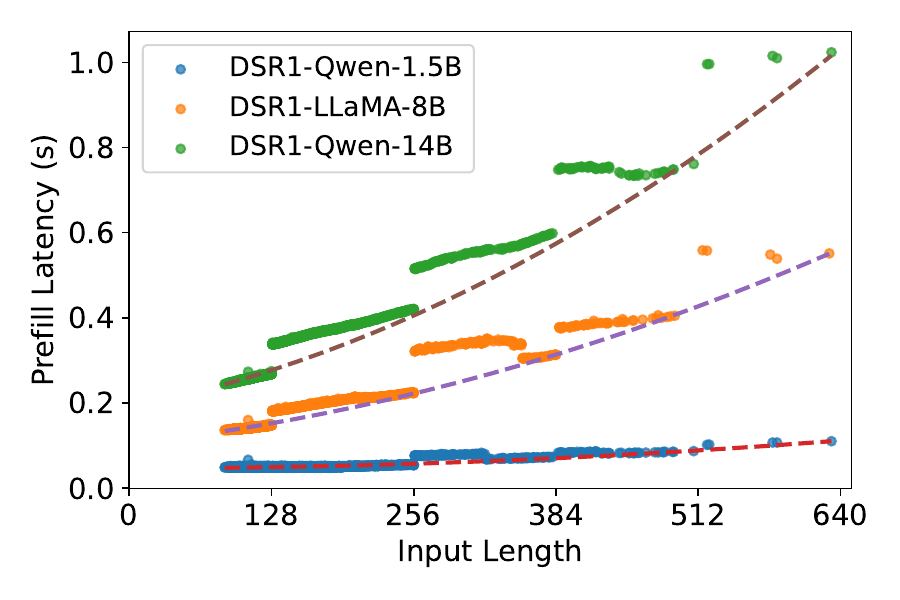}
    \caption{ Prefill latency vs. input sequence length.}
    \label{fig:prefill-time}
\end{figure}

\textbf{Prefill Latency.} Fig.~\ref{fig:prefill-time} illustrates how measured prefill latency varies with input token counts for single-batch inference.
% \jenny{TODO for JENNY might need a theoretical perf line vs the measured one}
% The measured data deviates significantly from the theoretical optimal latency calculated using the Orin roofline. 
We observe a distinctive stepped pattern where latency exhibits sub-quadratic scaling at token counts that are multiples of 128, with either linear increases or plateau regions within each 128-token segment. 

This behavior stems from tensor quantization effects in the CUTLASS kernels when utilizing Tensor Cores, which process data in fixed-size blocks and require padding to align with hardware-optimized dimensions.
The performance characteristics within these 128-token segments depend on computational intensity. At lower token counts, the system operates in a memory-bandwidth-limited region where kernels are constrained by memory bandwidth rather than compute capacity, resulting in linear latency increases over input token counts within each segment. As prefill token counts grow larger, the system transitions to a compute-bound region where the kernels become limited by arithmetic throughput. In this regime, padding effects become more pronounced since workloads within the same token segment require identical FLOPS, leading to the observed plateau behavior. Additional performance variations that deviate from the primary trend are likely attributable to the selection of different CUTLASS kernel variants optimized for different GEMM shapes. 

For the prefill phase of a given LLM, the theoretical compute and memory complexity scales linearly with input length \( I \) in the projection and feedforward layers, and quadratically in the attention layers. Based on this, we model the prefill latency as a quadratic function: $L_{\text{prefill}}(I) = aI^2 + bI + c$. 

To account for Tensor Core padding effects, we restrict the model fitting to data points where the input length is a multiple of 64. In practice, input lengths are rounded up to the nearest multiple of 128 to form a padded length $I_{\text{pad}}$, defined as $I_{\text{pad}} = \left\lceil \frac{I}{128} \right\rceil \cdot 128$

Substituting \( I \) with \( I_{\text{pad}} \), the fitted prefill latency functions can be expressed as:
\begin{align}
L_{\text{prefill}}(I) = aI_{\text{pad}}^2 + bI_{\text{pad}} + c
\label{eqn:latency_prefill}
\end{align}

Table~\ref{tab:latency_prefill} lists the fitted coefficients for the prefill latency models of the 1.5B, 8B, and 14B DSR1 models. The fitted functions are also plotted in dashed lines in Fig.~\ref{fig:prefill-time}. 

\begin{table}[t]
\centering
\caption{Fitted coefficients for prefill latency model.}
% : \( L_{\text{prefill}} = a \cdot I_{\text{pad}}^2 + b \cdot I_{\text{pad}} + c \)}
\label{tab:latency_prefill}
\begin{tabular}{lccc}
\toprule
\textbf{Model} & \boldmath$a$ & \boldmath$b$ & \boldmath$c$ \\
\midrule
DSR1-Qwen-1.5B   & \( 1.56 \times 10^{-7} \) & \( 2.31 \times 10^{-6} \) & \( 0.046 \) \\
DSR1-LLama-8B    & \( 6.65 \times 10^{-7} \) & \( 2.90 \times 10^{-4} \) & \( 0.104 \) \\
DSR1-Qwen-14B    & \( 1.23 \times 10^{-6} \) & \( 5.3 \times 10^{-4} \) & \( 0.189 \) \\
\bottomrule
\end{tabular}
\end{table}

% {\small
% \begin{align}
% \text{Qwen-1.5B:} \quad & L_{\text{prefill}} = 9.0 \times 10^{-8} \cdot I_{\text{pad}}^2 + 5.1 \times 10^{-5} \cdot I_{\text{pad}} + 0.041 \\
% \text{LLaMA-8B:} \quad & L_{\text{prefill}} = 2.2 \times 10^{-6} \cdot I_{\text{pad}}^2 - 4.0 \times 10^{-4} \cdot I_{\text{pad}} + 0.16 \\
% \text{Qwen-14B:} \quad & L_{\text{prefill}} = 1.2 \times 10^{-6} \cdot I_{\text{pad}}^2 + 5.3 \times 10^{-4} \cdot I_{\text{pad}} + 0.19
% \end{align}
% }

% \begin{figure}[t]
%     \centering
%     \includegraphics[width=0.80\linewidth]{figs/decode_latency/all_models_decode_latency_rect.pdf}
%     \caption{Decoding latency vs. output length with fixed input length.
%     }
%     \label{fig:decode-time}
% \end{figure}
% \begin{figure}[t]
%     \centering
%     \includegraphics[width=0.85\linewidth]{figs/llama8b_decode_latency.pdf}
%         \caption{Decoding latency vs. output length with varying input lengths for DSR1-Llama-8B.}
%     \label{fig:decode-time8B}
% \end{figure}

\begin{table}[t]
\centering
\caption{Fitted coefficients for decode latency model}
\label{tab:latency_decode}
\begin{tabular}{lcc}
\toprule
\textbf{Model} & \boldmath$m$ & \boldmath$n$  \\
\midrule
DSR1-Qwen-1.5B   & \( -1.50 \times 10^{-7} \) & $0.024$ \\
DSR1-LLama-8B    & $6.92 \times 10^{-7}$  & $0.010$ \\
DSR1-Qwen-14B    & $1.13 \times 10^{-6}$ &  $0.187$ \\
\bottomrule
\end{tabular}
\end{table}

\begin{figure}[t]
   \begin{subfigure}{0.48\linewidth}
      \centering
      % left bottom right top
      \includegraphics[trim=0 0 0 0, clip, width=\linewidth]{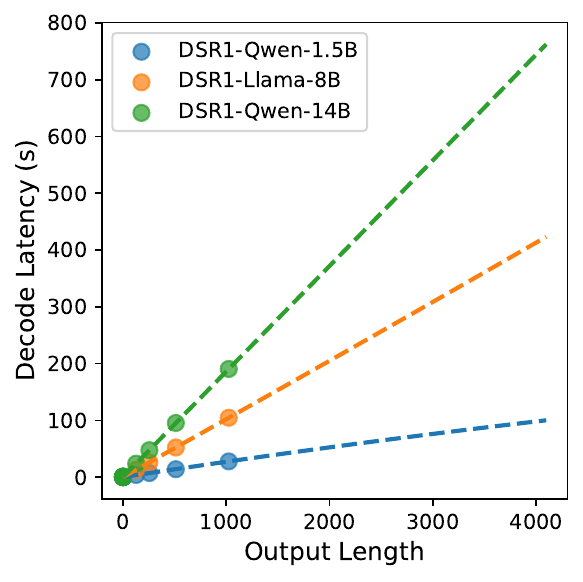}
      \caption{Input length=512.}
    \label{fig:latency_decode_fixed_input}
  \end{subfigure}
  \hfill
     \begin{subfigure}{0.48\linewidth}
      \centering
      % left bottom right top
      \includegraphics[trim=0 0 0 0, clip, width=\linewidth]{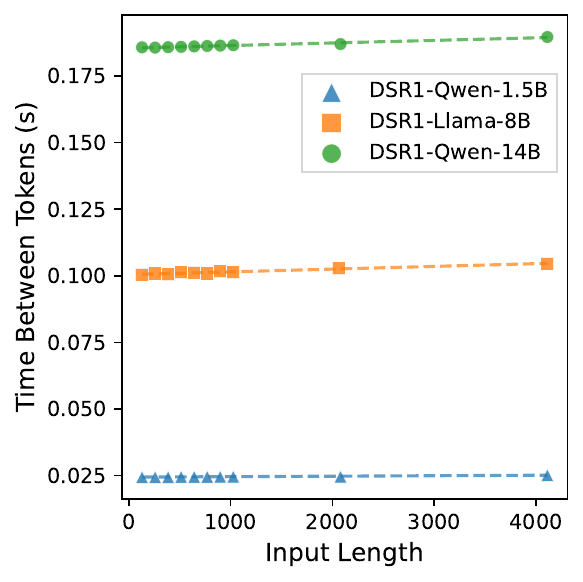}
      \caption{Time between tokens}
      \label{fig:latency_decode_tbt}
      \end{subfigure}
      % \vspace{-0.3cm}
      \caption{Decode latency vs output and input sequence lengths.}\label{fig:decode_latency}

\end{figure}

\textbf{Decode Latency.}
Fig.~\ref{fig:latency_decode_fixed_input} demonstrates how measured decode latency varies with output length $O$ with fixed input length of 512 for three different models. Decode latency always grows near linearly with respect to the output length due to the autoregressive nature of the decoding process. 
Fig.~\ref{fig:latency_decode_tbt} shows how the time between tokens varies with input length $I$ for the DSR1-Llama-8B model. We observe a slight 3.1\% TBT increase with input length increases from 1 to 4k.  

At each decoding step $i$, the input context length increases by one, i.e., $I_i = I_{i-1} + 1$. Since the attention layer's latency grows linearly with the input context length during decode, we model the time between tokens (TBT) as $TBT_i = mI_i + n$. 
The total decoding latency, $L_{\text{decode}}$, is the sum of TBT across all $O$ output steps: $L_{\text{decode}} = \sum_{i=0}^{O-1} TBT_i$. 
By simplifying the expression for $L_{\text{decode}}$, we obtain the following theoretical decode latency model:

\begin{align}
L_{\text{decode}}(I, O) = nO + m \left( IO + \frac{O(O - 1)}{2} \right)
\label{eqn:latency_decode}
\end{align}
% https://chatgpt.com/s/t_685fc4fea3e481918603248694b51c19 
where \( I \) is the initial input length. 

To derive the decode latency model, we first fit the decode latency model in Eqn.~\ref{eqn:latency_decode} using 100 MMLU-Redux data points with various input and output lengths. The corresponding coefficients $m$ and $n$ for different LLM are listed on Table~\ref{tab:latency_decode}.

Since the $m$ is very small, the TBT is almost equal to $n$. 
The average time between tokens (TBT) for the 1.5B, 8B, and 14B models are 0.029s, 0.092s, and 0.187s, respectively. They are corresponding to the slopes of lines in Fig.~\ref{fig:latency_decode_tbt}.

Given the negligible magnitude of the slope coefficient $m$, the average TBT can be effectively approximated by $n$. The corresponding TBT values for the 1.5B, 8B, and 14B models are 0.024s, 0.10s, and 0.186s, respectively. These values correspond to the slopes of the curves shown in Fig.~\ref{fig:latency_decode_tbt}, confirming that TBT remains relatively constant across different context lengths for each model size.

\textbf{Total latency.} Combining Eqn.~\ref{eqn:latency_prefill} and ~\ref{eqn:latency_decode}, we have the total inference latency on Jetson Orin GPU for the three models defined as: 
\begin{align}
    L = L_{\text{prefill}} + L_{\text{decode}} 
\label{eqn:total_latency}
\end{align}

We validate our fitted analytical latency models on 50 held-out MMLU-Redux test questions. Table~\ref{tab:latency_mape_analysis} shows that the predicted latencies match the measured values closely, with total MAPE under 2\% across all models.
We use these fitted latency models throughout the remainder of this paper to accelerate latency evaluation and optimal inference strategy search, as real measurements on the complete dataset to produce one latency point would require weeks to finish. For instance, a full latency evaluation on all MMLU-Redux questions using DSR1-LLaMA-14B takes 8 days to complete, while the analytical model produces results within seconds.

\begin{table}[H]
\centering
\caption{Mean Absolute Percentage Error (MAPE) of Latency Model}
\label{tab:latency_mape_analysis}
\begin{tabular}{lccc}
\toprule
\textbf{Model} & \textbf{Prefill} & \textbf{Decode} & \textbf{Total} \\
\midrule
DSR1-Qwen-1.5B   & 9.80\%  & 0.42\%  & 0.46\% \\
DSR1-LLaMA-8B    & 13.39\% & 0.45\%  & 0.49\% \\
DSR1-Qwen-14B    & 7.59\% & 0.53\% & 0.56\%\\
\bottomrule
\end{tabular}
\end{table} 
\begin{keytakeaway}
\textbf{Takeaway \#1:} Edge inference latency of LLMs can be accurately fitted using polynomial functions.
\end{keytakeaway}

\textbf{Prefill-to-decode latency ratio for reasoning models.}
Table~\ref{tab:token_latency_ratios} presents the prefill-to-decode token and latency ratios when running the complete MMLU-Redux dataset across our three reasoning models. The results reveal a striking disparity between token generation patterns and actual latency distribution. While the models generate 2.4-7.3× more decode tokens than prefill tokens, the latency imbalance is far more pronounced, with decode phase consuming 192-569× longer than prefill phase. This dramatic difference stems from the sequential nature of autoregressive generation during decode, where each token must be generated individually, compared to the parallel processing of all input tokens during prefill. The Qwen models exhibit higher token ratios (7.1-7.3×) due to their more verbose reasoning chains, yet all models show consistently extreme latency ratios, with decode dominating over 99.5\% of total inference time. This analysis underscores the critical importance of decode optimization for reasoning workloads on edge devices. 
\begin{table}[H]
\centering
\caption{Prefill-to-decode Ratios for Full MMLU-Redux}
\label{tab:token_latency_ratios}
\begin{tabular}{lcc}
\toprule
\textbf{Model} & \textbf{P-to-D Tokens Ratio} & \textbf{P-to-D Latency Ratio} \\
\midrule
DSR1-Qwen-1.5B & 1:7.3 & 1:521 \\
DSR1-LLaMA-8B  & 1:2.4 & 1:192 \\
DSR1-Qwen-14B  & 1:7.1 & 1:569 \\
\bottomrule
\end{tabular}
\end{table}

\begin{keytakeaway}
\textbf{Takeaway \#2:} Edge inference latency of reasoning LLMs is dominated by decode.
\end{keytakeaway}

% The points on the plot represent empirical measurements from runs while the dashed lines depict fitted linear performance models for the three different model sizes. 
 
% \subsubsection{Sequence Length}
% \todo{
% include the following figures
% }
% % \begin{figure}[H]
% %     \centering
% %     \includegraphics[width=0.9\linewidth]{figs/fig2/combined_prefill_fit_allpoints_closest.pdf}
% %     \caption{ OLD Prefill latency vs. input length.}
% %     \label{fig:prefill-time}
% % \end{figure}

% \begin{enumerate}
%     \item prefill token counts (x-axis) vs prefill latency (y-axis) for some MMLU-Redux questions for 3 models in 1.5, 8B, 14B (different color points). fit an analytical model and plot them in 3 solid color lines.
%     \item decoding only token counts (x-axis) vs decoding only latency (y-axis) assuming fixed 512 prefilling context for 3 models in 1.5, 8B, 14B (different color points). 
%     \item decoding only token counts (x-axis) vs decoding latency (y-axis) for 8B model, with prefilling context in [1, 128, 256, 512]
    
%     \item fits a power function
%     \todo{take the power number of fig 2}
%     % \item Decoding latency ($Latency_{decoding} = f(Length_{prefilling}, Length_{decoding})$)  is a function of prefilling length and decoding length. after fitting the function, plot a 3D figure with x axis = $Length_{prefilling}$, y=$Length_{decoding})$, z=$Latency_{decoding}$ for 8B model.  
% \end{enumerate}

\subsection{Characterization of Power and Energy}

% https://docs.nvidia.com/jetson/archives/r35.1/DeveloperGuide/text/SD/PlatformPowerAndPerformance/JetsonOrinNxSeriesAndJetsonAgxOrinSeries.html
Beyond latency, understanding power consumption and energy efficiency is critical for edge deployment scenarios. The Jetson AGX Orin 64GB platform we used in the study supports four configurable power modes (15W, 30W, 50W, and MAXN) that set peak frequencies across GPU, CPU, DLA, and PVA units. All experiments are conducted in MAXN mode to capture peak performance characteristics. We analyze power consumption and energy usage as functions of input length, output length, and model size to establish fundamental scaling relationships for edge inference workloads.

% \begin{figure}[t]
%     \centering
%     \includegraphics[width=0.7\linewidth]{figs/prefill_energy/power_fits_base_reasoning.pdf}
%     \caption{Prefill power vs. input sequence length.}
%     \label{fig:power_prefill}
% \end{figure}

% \begin{figure}[t]
%     \centering
%     \includegraphics[width=0.7\linewidth]{figs/prefill_energy/energy_per_token_fits.pdf}
%     \caption{Energy per token vs input length.}
%     \label{fig:energy_prefill}
% \end{figure}

\begin{figure}[t]
    \centering
    %--- Prefill power --------------------------------------------------
    \begin{subfigure}[b]{0.48\linewidth}
        \centering
        \includegraphics[width=\linewidth]{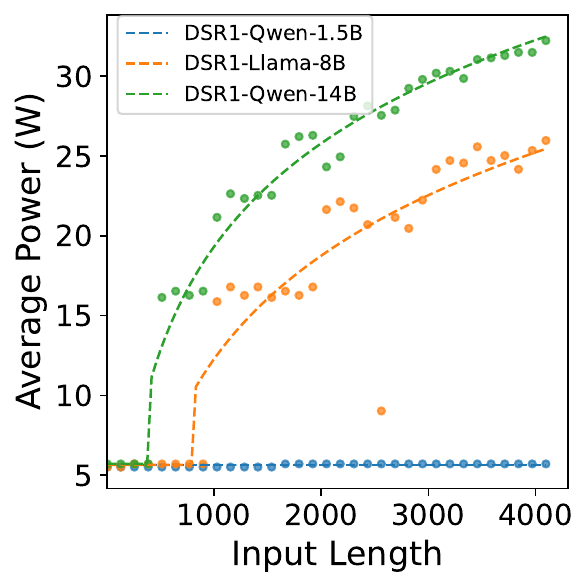}
        \caption{Prefill power.}
        \label{fig:power_prefill}
    \end{subfigure}
    \hfill
    %--- Energy per token ----------------------------------------------
    \begin{subfigure}[b]{0.48\linewidth}
        \centering
        \includegraphics[width=\linewidth]{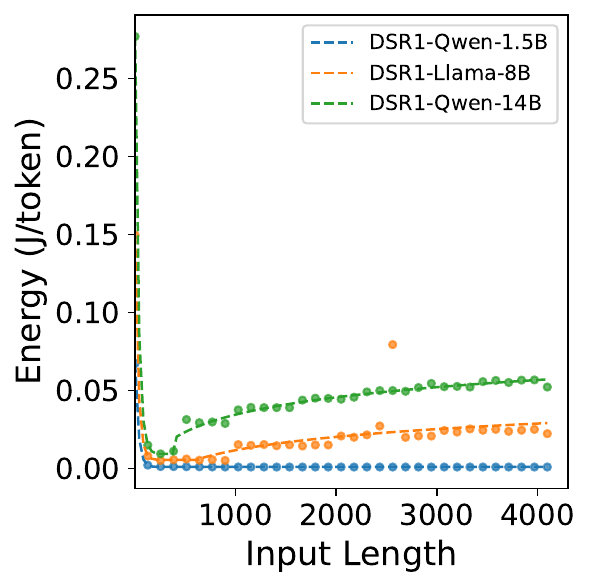}
        \caption{Energy per token.}
        \label{fig:energy_prefill}
    \end{subfigure}

    \caption{Prefill power (left) and energy per token (right) as a function of input sequence length.}
    \label{fig:prefill_energy_overview}
\end{figure}

% \begin{figure}[t]
%     \centering
%     \includegraphics[width=0.7\linewidth]{figs/decode_energy/power_consumption_fits.pdf}
%     \caption{Decode power vs. output sequence length.}
%     \label{fig:power_decode}
% \end{figure}

% \begin{figure}[t]
%     \centering
%     \includegraphics[width=0.7\linewidth]{figs/decode_energy/energy_token_fits.pdf}
%     \caption{Decode per token vs output sequence length.} %\todo{energy / tps is more interesting}}
%     \label{fig:energy_decode}
% \end{figure}

\begin{figure}[t]
    \centering
    %--- Decode power ---------------------------------------------------
    \begin{subfigure}[b]{0.48\linewidth}
        \centering
        \includegraphics[width=\linewidth]{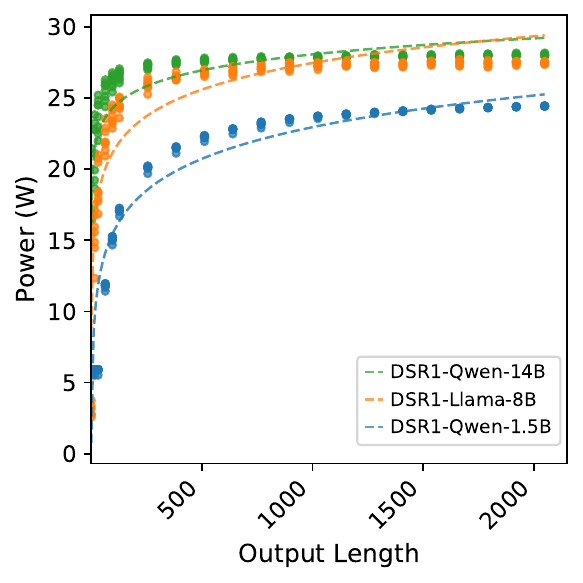}
        \caption{Power.}
        \label{fig:power_decode}
    \end{subfigure}
    \hfill
    %--- Energy per token ----------------------------------------------
    \begin{subfigure}[b]{0.48\linewidth}
        \centering
        \includegraphics[width=\linewidth]{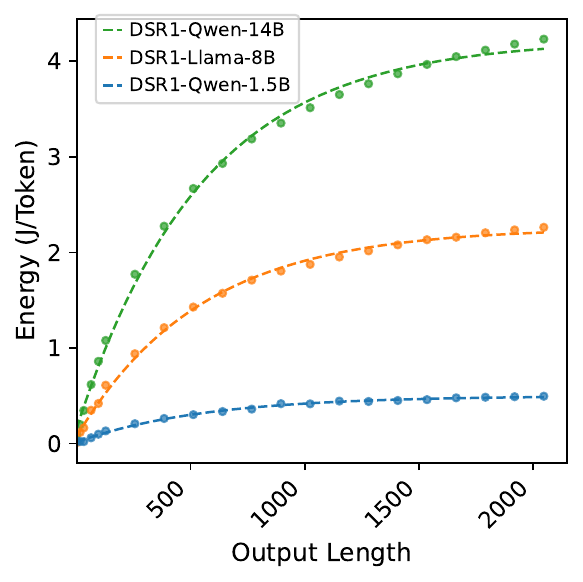}
        \caption{Energy per token.}
        \label{fig:energy_decode}
    \end{subfigure}

    \caption{Decode phase power (left) and energy per token (right) as a function of output sequence length}
    \label{fig:decode_energy_overview}
\end{figure}

% \begin{figure}[t]
%     \centering
%     \includegraphics[width=0.75\linewidth]{figs/ps_orin/decode_latency_vs_ps.pdf}
%     \caption{Decode latency vs. parallel scaling factor.}
%     \label{fig:latency_decode_ps}
% \end{figure}
% \begin{figure}[t]
%     \centering
%     \caption{Power vs. parallel scaling factor.}
%     \label{fig:power_gpu_decode_ps}
% \end{figure}
% \begin{figure}[t]
%     \centering
%     \includegraphics[width=0.75\linewidth]{figs/ps_orin/energy_per_decode_vs_ps.pdf}
%     \caption{Energy per question vs. parallel scaling factor.}
%     \label{fig:energy_decode_ps}
% \end{figure}

\textbf{Prefill Power and Energy.} 
Fig.~\ref{fig:power_prefill} shows that average power consumption increases with input sequence length for single-batch inference, measured on the Jetson AGX Orin with 5 repeated samples per data point. This trend occurs because longer input sequences increase the computational intensity of the workload, leading to higher GPU utilization.
The larger 8B and 14B models reach over 20W at 4K input sequence length, while the smaller 1.5B model consumes only 6W—representing just 10\% of the platform's 60W peak power capacity. This shows the significant differences in power consumption between model sizes.
 
Fig.~\ref{fig:energy_prefill} shows the energy consumption per input token across different input sequence lengths. The results demonstrate that smaller models consistently achieve superior energy efficiency compared to larger models due to their reduced FLOPs and memory requirements.
Across all three models, we observe a characteristic trend where energy per token initially decreases from short input lengths until reaching a minimum around 300 tokens. We attribute this behavior to the dominance of projection and feed-forward (FFN) layers in this regime, where increased input length leads to better weight reuse and improved energy efficiency.
Beyond this point, as the workload becomes attention-bound, further increases in input length provide diminishing returns from weight reuse. Consequently, we observe that energy per token plateaus for large input sequences, with oscillations around the steady-state value.

We also develop an analytical power model based on the observed data to accelerate evaluation. Since the prefill power consumption $P_{\text{prefill}}(I)$ exhibits two distinct regimes depending on input length $I$, we define our power model as:
\begin{align}
P_{\text{prefill}}(I) =
\begin{cases}
u, & I \le v,\\[6pt]
w\ln(I) + x, & I > v.
\end{cases}
\label{eqn:ppwer_prefill}
\end{align}
For shorter input sequences ($I \le v$), power remains constant at $u$ watts, indicating low GPU utilization. For longer sequences ($I > v$), power consumption increases logarithmically, reflecting higher computational intensity and improved hardware utilization.
The implied energy consumption model follows from $
E_{\text{prefill}}(t) = \int_0^{L_{\text{prefill}}(I)} P_{prefill}(t)\,dt$, where energy is the time integral of instantaneous power. We additionally fit a direct piecewise energy model that captures the
amortization of short-sequence overheads and energy increase at longer lengths:

\begin{equation}
E_{\text{prefill}}(I)=
\begin{cases}
A\,e^{-\lambda I}+C, & I \le v_e,\\[2pt]
\alpha_e \ln I + \beta_e, & I > v_e,
\end{cases}
\label{eqn:energy_prefill_base}
\end{equation}
where \(v\) and \(v_e\) are model-specific transition points. For the distilled models used here,
typical transitions are \(v{=}800\) (8B) and \(v{=}384\) (14B), while the 1.5B case is effectively
constant over the measured range. The fitted coefficients
for each model are provided in appendix Table~\ref{tab:prefill_fit_params_base}.

% MAPE of Energy Model
\begin{table}[H]
  \centering
  \caption{Mean Absolute Percentage Error (MAPE) of Energy Model}
  \label{tab:energy_mape_analysis}
  \begin{tabular}{lccc}
    \toprule
    \textbf{Model} & \textbf{Prefill} & \textbf{Decode} & \textbf{Total} \\
    \midrule
    DSR1-Qwen-1.5B  & -- & 6.8\% & 6.0\% \\
    DSR1-Llama-8B   & -- & 6.4\% & 5.7\% \\
    DSR1-Qwen-14B   & -- & 6.6\% & 5.8\% \\
    \bottomrule
  \end{tabular}
\end{table}

\textbf{Decode Power and Energy.} 
Fig.~\ref{fig:power_decode} shows the average power consumption and energy for varying output sequence lengths with a fixed input sequence length of 512 tokens. The results demonstrate that power consumption increases logarithmically with output sequence length. While the computation of the projection and FFN layers remains constant during decoding, this increase is due to the growing computational and memory demands in the attention layer as the context window expands.
The analysis also demonstrates significant efficiency gains from model size reduction: the 1.5B model achieves a 7× improvement in energy per token compared to the 14B model, highlighting the substantial energy benefits of deploying smaller models for resource-constrained edge environments. 
The decode power is fitted with the same functional form used for the prefill power in Eqn.~\ref{eqn:ppwer_prefill}
For the orin GPU, this yields Eqn.~\ref{eqn:power_decode}.

\begin{align}
P_{\text{decode}}(O) =
\begin{cases}
5.9\,\mathrm{W}, & 0 < O < 64,\\[4pt]
y\ln O + z, & O \ge 64~,
\end{cases}
\label{eqn:power_decode}
\end{align}

where $O$ is the output sequence length, and $y$ and $z$ are fitted parameters that capture the logarithmic scaling behavior observed in our measurements of different models.
The corresponding energy consumption is given by $E_{\text{decode}}(t) = \int_0^{L_{\text{decode}}(I)} P_{\text{decode}}(t)\,dt$. 

\textbf{Total Energy.} 
We model the total energy as $E = E_{\text{prefill}}(O) + E_{\text{decode}}(O)$. 
Since decode latency is significantly longer than prefill, decode energy consumption also dominates the total energy budget.

\begin{keytakeaway}
\textbf{Takeaway \#3:} Average power and total energy consumption increase logarithmically with sequence length on NVIDIA Jetson AGX Orin platform.
\end{keytakeaway}

% include the following figures
% \begin{enumerate}
% \item accuracy vs token length (full MMLU-Redux)
% \item accuracy vs inference latency (full MMLU-Redux)
% \item accuracy vs energy/cost (full MMLU-Redux) assuming energy can be extrapolated from power x latency, calculate both energy and energy cost in x axis. 
% \item plot bar charts for average GPU utilization(even better if we have collected sm, tensor core, mem util) for ps=1, 128to1, 1to8 decoding data for 3 models. (3x2 points)
% \end{enumerate}
% Reasoning Token Control Methods

% \textbf{Comparison of various token control techniques}.
\section{ Evaluation of Inference Strategies}

% \begin{itemize}
%   \item \textcolor{col128T}{\textbf{$\times$}} 128T
%   \item \textcolor{col128TNC}{$\Diamond$} 128T-NC
%   \item \textcolor{col256T}{$\triangle$} 256T
%   \item \textcolor{col256TNC}{$\triangledown$} 256T-NC
%   \item \textcolor{colBase}{$\circ$} Base
%   \item \textcolor{colNR}{$\star$} NR
%   \item \textcolor{colDirect}{$+$} Direct
%   \end{itemize}

This section compares different inference strategies for showing the accuracy-latency/energy tradeoffs and guiding optimal inference strategies on edge GPUs. 
We evaluate the tradeoffs for reasoning vs non-reasoning models and reasoning models in different sizes.  We also evaluate different prompt-based and fine-tuning-based methods for reducing output sequence length while maintaining accuracy. 

This section systematically evaluates inference strategies to quantify accuracy-latency and cost tradeoffs and guide optimal configuration selection for reasoning models on edge GPUs. We analyze three critical dimensions: (1) reasoning capability vs. model size tradeoffs, (2) output sequence length reduction techniques, and (3) energy and cost efficiency implications. Our evaluation encompasses three model categories:
\begin{itemize}
    \item \textbf{Standard Models (Non-reasoning):} Baseline architectures generating direct responses without explicit reasoning chains: \emph{Qwen2.5-1.5B-it}~\cite{qwen2.5}, \emph{Llama3.1-8B-it}~\cite{grattafiori2024llama}, and \emph{Qwen2.5-7B-it}~\cite{qwen2.5}.
    
    \item \textbf{Reasoning Models:} Lightweight reasoning-optimized models from the Deepseek-R1 (DSR1) family: \emph{DSR1-Qwen-1.5B}, \emph{DSR1-Llama-8B}, and \emph{DSR1-Qwen-14B}. These distilled models perform standard autoregressive inference without token constraints.
    
    \item \textbf{Budget-Aware Reasoning Model:} \emph{L1}~\cite{aggarwal2025l1}, a \emph{DSR1-Qwen-1.5B} variant fine-tuned via reinforcement learning to maximize accuracy under user-specified token budgets.
\end{itemize}

For output length optimization, we evaluate three prompt-based approaches applied to reasoning models:

\begin{itemize}
    \item \textbf{Hard-Length Control ([n]T):} Explicit length instructions (e.g., "Answer in [n] words") with strict token enforcement. Configurations: \emph{128T}, \emph{256T}.
    
    \item \textbf{Soft-Length Control ([n]-NC):} Identical instructions without token enforcement. Configurations: \emph{128-NC}, \emph{256-NC}.
    
    \item \textbf{No Reasoning (NR):} Bypasses explicit reasoning by injecting predefined thinking blocks between delimiters~\cite{ma2025reasoning}:
    \begin{tcolorbox}[colback=white!10,colframe=gray!50,arc=0pt,boxrule=0.5pt]
    \footnotesize
    \texttt{<|beginning of thinking|>} \\
    \texttt{Okay, I think I have finished thinking.} \\
    \texttt{<|end of thinking|>}
    \end{tcolorbox}
\end{itemize}

\begin{figure}[t]
    \centering
    \begin{subfigure}[b]{0.325\linewidth}
        \centering
        \includegraphics[width=\linewidth]{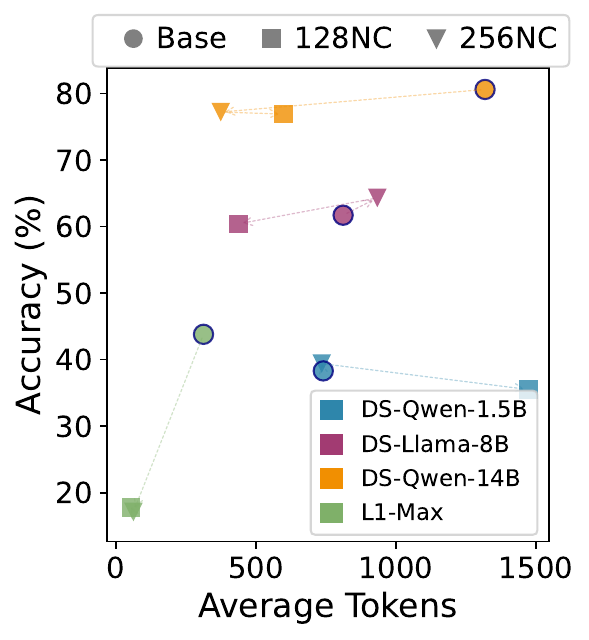}
        \caption{Soft limit}
        \label{fig:soft_tokens}
    \end{subfigure}
    \hfill
    \begin{subfigure}[b]{0.325\linewidth}
        \centering
        \includegraphics[width=\linewidth]{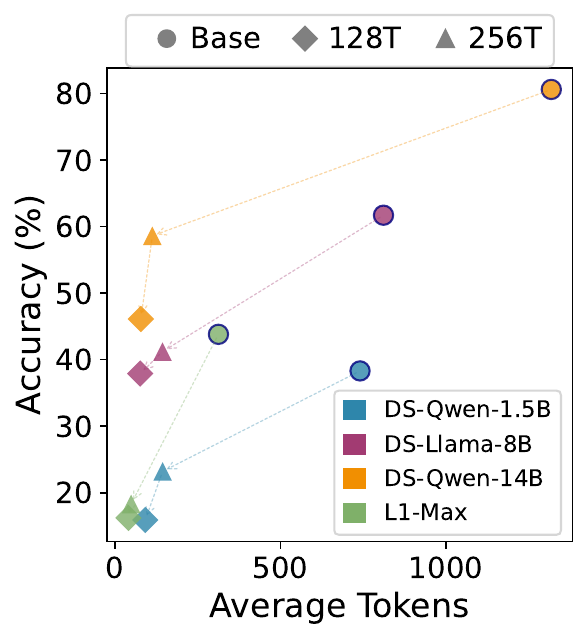}
        \caption{Hard limit}
        \label{fig:hard_tokens}
    \end{subfigure}
    \hfill
    \begin{subfigure}[b]{0.325\linewidth}
        \centering
        \includegraphics[width=\linewidth]{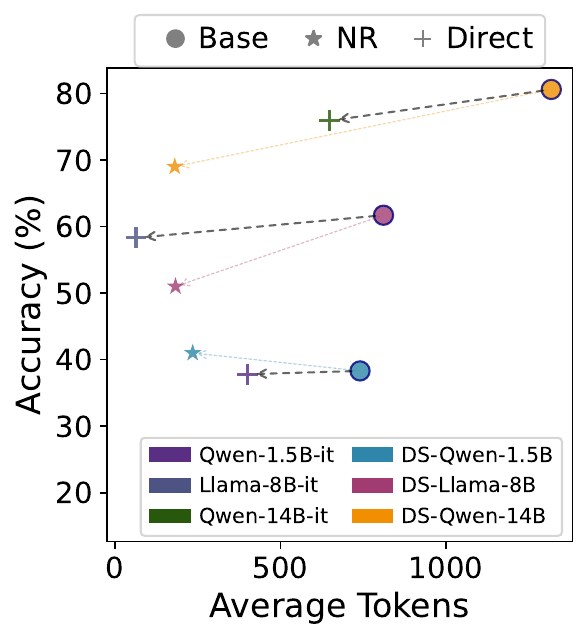}
        \caption{No reasoning}
        \label{fig:nr_tokens}
    \end{subfigure}
    \caption{Accuracy versus average output length across budgeting techniques.}
    \label{fig:budget_overview}
\end{figure}

\begin{figure*}[t]
    \centering
    \begin{subfigure}[b]{0.28\linewidth}
        \centering
        \includegraphics[width=\linewidth]{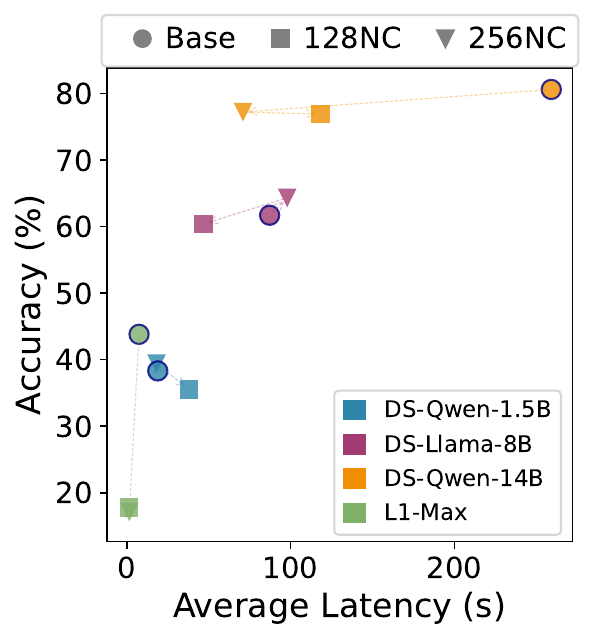}
        \caption{Soft limit}
        \label{fig:soft_latency}
    \end{subfigure}
    \hfill
    \begin{subfigure}[b]{0.28\linewidth}
        \centering
        \includegraphics[width=\linewidth]{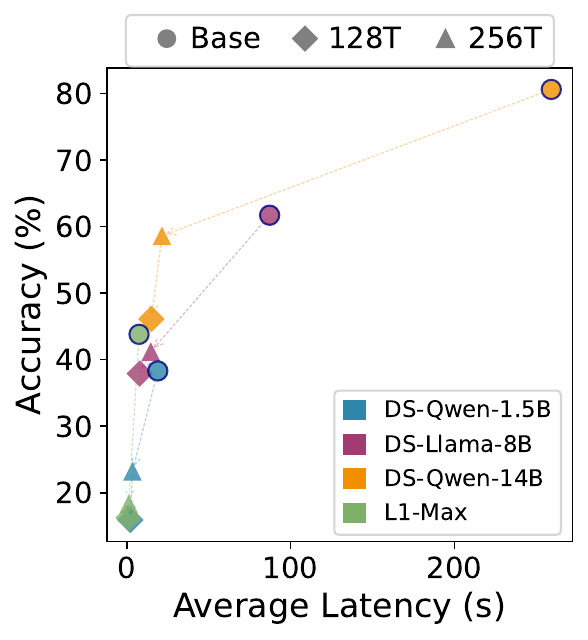}
        \caption{Hard limit}
        \label{fig:hard_latency}
    \end{subfigure}
    \hfill
    \begin{subfigure}[b]{0.28\linewidth}
        \centering
        \includegraphics[width=\linewidth]{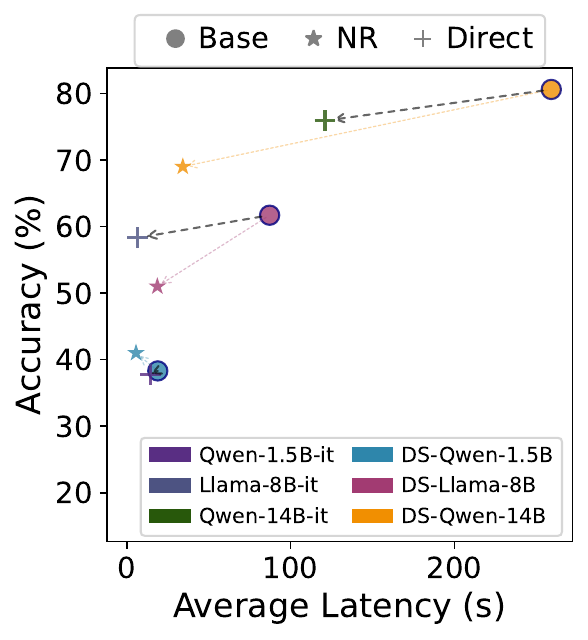}
        \caption{No reasoning}
        \label{fig:nr_latency}
    \end{subfigure}
    \caption{Accuracy versus latency across budgeting techniques.}
    \label{fig:latency_overview}
    \vspace{-0.5cm}

\end{figure*}

\begin{figure}[t]
    \centering
    \begin{subfigure}[b]{0.325\linewidth}
        \centering
        \includegraphics[width=\linewidth]{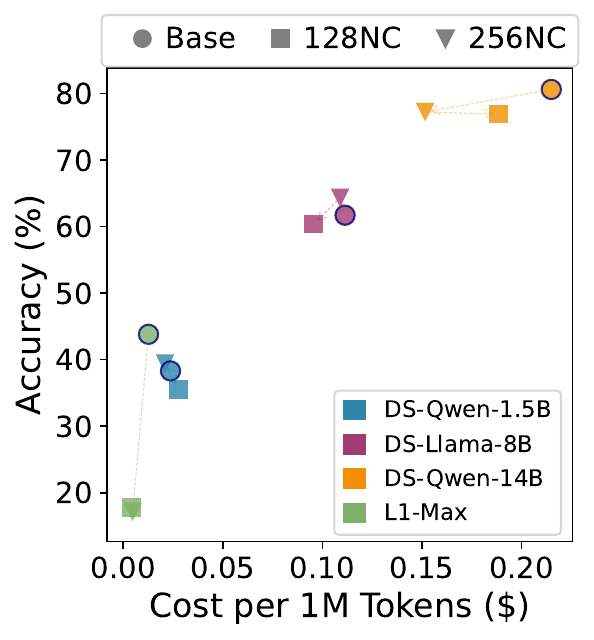}
        \caption{Soft limit}
        \label{fig:soft_cost}
    \end{subfigure}
    \hfill
    \begin{subfigure}[b]{0.325\linewidth}
        \centering
        \includegraphics[width=\linewidth]{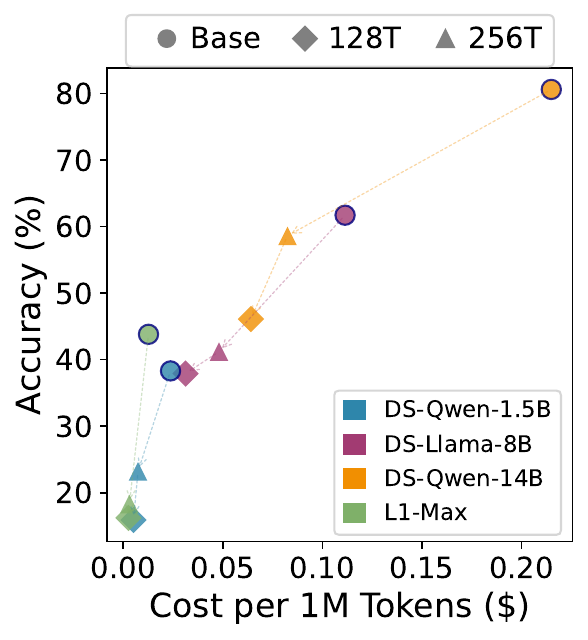}
        \caption{Hard limit}
        \label{fig:hard_cost}
    \end{subfigure}
    \hfill
    \begin{subfigure}[b]{0.325\linewidth}
        \centering
        \includegraphics[width=\linewidth]{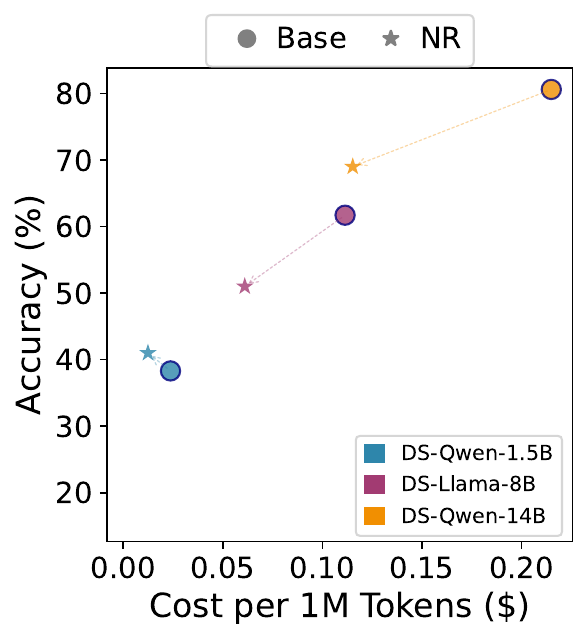}
        \caption{No reasoning}
        \label{fig:nr_cost}
    \end{subfigure}
    \caption{Accuracy versus cost across budgeting techniques.}
    \label{fig:cost_overview}
    \vspace{-0.5cm}
\end{figure}

%Still missing the 14B model data for both direct and quantized. I will add the plots when the data is ready.

All configurations are evaluated on the MMLU-Redux benchmark~\cite{gema2024mmlu}, comprising 3,000 multiple-choice questions spanning humanities, social sciences, STEM, and professional domains. The benchmark tests both factual knowledge and reasoning capabilities across difficulty levels from elementary to graduate. 
For each configuration, we report four key performance metrics:
(1) accuracy on the MMLU-Redux benchmark, 
(2) average decoded tokens per question, 
(3) average inference latency per question, and
(4) average cost per million tokens derived from energy measurements.

Our results reveal fundamental tradeoffs between critical metrics:
\begin{itemize}
    \item \textbf{Accuracy vs. Output Length:} Fig.~\ref{fig:budget_overview} demonstrates how accuracy varies with generated sequence length across model classes and length-control methods, revealing the compression-performance frontier.
    
    \item \textbf{Accuracy vs. Latency:} Fig.~\ref{fig:latency_overview} quantifies the accuracy-latency tradeoff on edge hardware.

    \item \textbf{Accuracy vs. Cost Efficiency:} Fig.~\ref{fig:cost_overview} correlates reasoning quality with operational economics through cost per million tokens.
\end{itemize}

\subsection{Impact of Model Selection}
Model selection significantly impacts the achievable accuracy-latency tradeoff, making it crucial to understand how to choose between reasoning and non-reasoning architectures, as well as among reasoning models of different sizes.

% (e.g., Qwen2.5-8B for 5s$\sim$20s budget)
% (5s$\sim$20s)

\textbf{Impact of model size.} 
Analysis of reasoning models across different sizes—\emph{DSR1-Qwen-1.5B} (\textcolor{DSR1-Qwen-1.5B}{blue}), \emph{DSR1-Llama-8B} (\textcolor{DSR1-Llama-8B}{plum}), and \emph{DSR1-Qwen-14B} (\textcolor{DSR1-Qwen-14B}{orange})—shows that larger models typically achieve higher accuracy at the cost of increased inference latency. 
Fig.~\ref{fig:budget_overview} demonstrates that larger models in their Base configuration ($\circ$ markers) naturally generate more reasoning tokens when unconstrained. 

Fig.~\ref{fig:budget_overview} reveals an intriguing trade-off space where smaller models with higher token budgets can be competitive with larger models operating under smaller token budgets. Notably, in Fig.~\ref{fig:hard_tokens}, \emph{DSR1-Llama-8B }\emph{Base}(\textcolor{DSR1-Llama-8B}{\symBase{}}) (generating 811 tokens on average) achieves higher accuracy than \emph{DSR1-Qwen-14B} 
\emph{128T}(\textcolor{DSR1-Qwen-14B}{\symonetwoeightT{}}) (generating only 91.5 tokens), suggesting that reasoning depth can compensate for reduced model scale.
Conversely, in Fig.~\ref{fig:soft_tokens}, \emph{DSR1-Llama-8B} \emph{Base}(\textcolor{DSR1-Llama-8B}{\symBase{}}) (generating 811 tokens on average) underperforms \emph{DSR1-Qwen-14B} \emph{256T-NC}(\textcolor{DSR1-Qwen-14B}{\symtwofivesixTNC{}})(generating only 374 tokens), suggesting that model scale can also compensate for reduced reasoning depth.
% From Fig.~\ref{fig:latency_overview}, we observed a tradeoff space between larger models with lower token budgets vs. smaller models with higher token budgets. For instance, the  \emph{DSR1-Llama-8B} \emph{Base}(\textcolor{DSR1-Llama-8B}{\symBase{}}) at token length of 811 underperforms the \emph{DSR1-Qwen-14B} \emph{256T-NC}(\textcolor{DSR1-Qwen-14B}{\symtwofivesixTNC{}}) at an average token counts of 374. 

% \emph{256T}(\textcolor{DSR1-Qwen-14B}{$\triangle$}) runs at 21s per question with 112.9 token counts, achieving similar accuracy to best \emph{DSR1-Llama-8B} \emph{Base}(\textcolor{DSR1-Llama-8B}{$\circ$}) data runs at 87s per question. Therefore, for a latency budget larger than 21s, \emph{DSR1-Qwen-14B} with a token length budget larger than 112 tokens should be preferred.
The crossover analysis provides practical deployment insights. 
For example, in Fig.~\ref{fig:hard_latency},  DSR1-Qwen-14B \emph{256T} (\textcolor{DSR1-Qwen-14B}{$\triangle$}) achieves comparable accuracy to DSR1-Llama-8B \emph{Base} at 4$\times$ lower latency (21s vs 87s) by operating within a 113-token budget. This indicates that for latency budgets exceeding 21s, DSR1-Qwen-14B with $>$113 token allocation becomes preferable.

% only 1.5B reasoning models provide viable solutions within 5-second latency constraints. For moderate latency budgets (15-30 seconds), the 8B non-reasoning models surprisingly outperform their reasoning counterparts. However, when latency budgets exceed 30 seconds, DSR1-Qwen-14B emerges as the optimal choice, offering the best accuracy-latency tradeoff for applications that can tolerate longer inference times. 

The Pareto-optimal frontier in Fig.~\ref{fig:latency_overview} reveals three distinct operational regimes:
\begin{itemize}
    \item \underline{Sub-5s latency:} Exclusively served by 1.5B models. 
    \item \underline{15-30s latency:} Non-reasoning 8B models are preferred.
    \item \underline{$>$30s latency:} DSR1-Qwen-14B emerges as optimal.
\end{itemize}

\begin{keytakeaway}
\textbf{Takeaway \#4:}  Only ultra-lightweight models (1.5B) can achieve real-time inference ($<$1s) on resource-constrained edge platforms.
\end{keytakeaway}
% Only ultra-lightweight models ($1.5B$) can achieve real-time inference ($<1s$) on NVIDIA AGX Orin platform.

\subsection{Impact of Reasoning Token Control Methods}
% Controlling output token length is critical for managing inference latency and meeting latency constraints. While the token-to-latency scaling factor varies per model, it can be characterized and fitted to an accurate performance model for NVIDIA Orin as demonstrated in Section~\ref{sec:latency}.
Output token length control is essential for latency management and meeting real-time constraints. While token-to-latency scaling factors are model-dependent, they exhibit predictable relationships that can be characterized and fitted to accurate performance models for the Orin GPU, as demonstrated in Section~\ref{sec:latency}.

% \textbf{Prompt-based approaches.}
% As shown in Fig.~\ref{fig:budget_overview}, adding a prompt to instruct model generate fewer tokens are effective in reducing inference latnecy, but oftentimes results in an accuracy drop as it minimizes sequential scaling for TTS.  We observed that in-prompt length control do not adhere to user specification. Take the \emph{128T-NC} configuration (marked as ``diamond'') for \emph{DSR1-Qwen-14B} (orange) as an example, it generates more than $4\times$  the token length than the \emph{128T} configuration with hard cutoff (marked as ``cross'') for \emph{DSR1-Qwen-14B} (orange). Nonetheless, prompting \emph{128T-NC} still makes the model token aware and reasoning process more efficient and generate $2\times$ fewer tokens than the uncontrolled generation by \emph{Base} but with comparable accuracy. 
\textbf{Prompt-based.}  
Fig.~\ref{fig:budget_overview} shows that in-prompt length control can significantly reduce output token length, but often at the cost of lower accuracy because it limits test-time scaling.
% the model’s effective token-time scaling. 
However, in-prompt length control rarely adheres to the user's specification.  For example, under the \emph{128-NC} setting (soft limit, \textcolor{DSR1-Qwen-14B}{\symonetwoeightTNC{}}) on \emph{DSR1-Qwen-14B}, the model emits four times as many tokens as the \emph{128T} (\textcolor{DSR1-Qwen-14B}{\symonetwoeightT{}}) with hard cutoff.  Even so, \emph{128-NC} still generates roughly half as many tokens as the uncontrolled \emph{Base} run, while maintaining comparable accuracy, demonstrating that the prompt makes the model modestly token-aware.

% The no-thinking approach \emph{NR} (marked as ``star'') provides another option to minimze output length and reduce reasoning time.  Comparing the \emph{NR} with \emph{Base} configuration in Fig.~\ref{fig:budget_overview}, \emph{NR} reduces the sequence length of \emph{DSR1-Qwen-1.5B} (blue) and \emph{DSR1-Qwen-14B} (orange), but increase the length of \emph{DSR1-Llama-8B} (plum). Compared \emph{NR} with \emph{Direct} on 8B models, it shows directly using a non-reasoning model achieves slightly higher accuracy performance with lower number of token generated. However,
% On \emph{DSR1-Qwen-1.5B} (blue) model, \emph{NR} achieves the highest accuracy, showing that turning off reasoning in small model can be effective. 

No-thinking \emph{NR} (marked as \symNR{}) provides another way to shorten outputs by skipping the explicit reasoning.  Compared to \emph{Base} (marked as \symBase{}), \emph{NR} reduces sequence length for all \emph{DSR1-Qwen-1.5B} (\textcolor{DSR1-Qwen-1.5B}{blue}), \emph{DSR1-Llama-8B} (\textcolor{DSR1-Llama-8B}{plum}), and \emph{DSR1-Qwen-14B} (\textcolor{DSR1-Qwen-14B}{orange}) models. When comparing \emph{DSR1-Llama-8B} \emph{NR} (\textcolor{DSR1-Llama-8B}{\symNR{}}) with the \emph{Direct} (marked as \symDirect{}) non-reasoning baselines, %(\textcolor{Qwen2.5-7B-it}{\symDirect{}}) 
 \emph{Direct} achieves slightly higher accuracy with fewer tokens, indicating that using a small non-reasoning model can outperform disabling reasoning in a larger one.  Interestingly, on the 1.5B model \emph{NR} attains the best accuracy overall, suggesting that suppressing the reasoning phase in very small models could be beneficial.

\begin{keytakeaway}
\textbf{Takeaway \#5:} Prompt-based approaches are effective in reducing reasoning tokens.
\end{keytakeaway}
\textbf{Budget-aware models.}
% As we show in Fig.~\ref{fig:budget_overview}, DSR1 family of models do not provide precision output length control, so we evaluated a new model \emph{L1} that is finetuned to achieve better instruction following capability on precision output length control. We use its \emph{L1-max} (purple) model for our inference, which requires the output legnth to be no longer than the token budget. 
% Figure~\ref{fig:budget_overview} shows that \emph{L1-max} without token budget limit (\emph{Base}, marked as dots) reaches higher accuracy than the its non-token-aware counterpart \emph{DSR1-Qwen-1.5B} (blue) while with more than 2$\times$ lower output token counts.  Meanwhile, the output seq length of \emph{L1-max} for in-prompt sequence length control configurations (\emph[n]T) meet the token length constraints. 
%  This demonstrates the effectiveness of RL finetuning in improving instruction capability on output length control.  
% However, we also observed it is overly conservative, generating lower than 50 tokens for a token budget of 256 in \emph{128T}. 
As shown in Fig.~\ref{fig:budget_overview}, standard DSR1 models lack precise output length control. To address this limitation, we evaluate \emph{L1}—a model specifically fine-tuned to enhance instruction-following capabilities for token budget adherence. Using its \emph{L1-max} variant (\textcolor{L1-Max}{green}), which strictly enforces output lengths within specified token budgets, we observe:
Without token constraints (\emph{Base}, \symBase{}), \emph{L1-max} achieves higher accuracy than \emph{DSR1-Qwen-1.5B} (\textcolor{DSR1-Qwen-1.5B}{blue}) while generating over 2× fewer tokens;
2) When constrained by in-prompt length specifications (e.g., \emph{128T-NC}(\symonetwoeightTNC{}), \emph{256T-NC}(\symtwofivesixTNC{}) in Fig.~\ref{fig:soft_tokens}), \emph{L1-max} consistently adheres to token budgets, demonstrating the efficacy of RL fine-tuning for output control.
However, we observe excessive conservatism: For a 256-token budget in the \emph{256T}(\symtwofivesixT) configuration, \emph{L1-max} generates fewer than 50 tokens—significantly underutilizing allocated capacity.
% Since \emph{L1} model effectively controls the token lengths, we can apply it in combination with an inverse function of the latency model we derive in Sec.~\ref{sec:latency}  Eqn.~\ref{eqn:total_latency} to derive an output token length constrains that map to latency constraint $L_C$. 
By leveraging the token length control capabilities of the \emph{L1} model and the analytical latency model from Eqn.~\ref{eqn:total_latency} (Sec.~\ref{sec:latency}), we can systematically determine output token length constraints that satisfy specified latency targets $L_C$.

\begin{keytakeaway}
\textbf{Takeaway \#6:} Fine-tuned token-budget-aware models combined with latency performance modeling enable adherence to latency constraints.
\end{keytakeaway}

\subsection{Impact of Sequential Test Time Scaling}
\label{sec:seq_scaling}
Fig.~\ref{fig:budget_overview} reveals a consistent trend across all base models: accuracy generally increases with output sequence length, regardless of the prompting approach. However, this relationship exhibits diminishing returns beyond certain token thresholds, 
specifically, $\sim$300 tokens for \emph{DSR1-Qwen-1.5B} (using L1's budget-aware tuning) and $\sim$400 tokens for both \emph{DSR1-Llama-8B} and \emph{DSR1-Qwen-14B}. 
These inflection points suggest where parallel scaling may surpass sequential scaling for accuracy gains, as shown in~\cite{muennighoff2025s1}.

The near-linear relationship between output length and inference latency (Section~\ref{sec:latency}, Fig.~\ref{fig:latency_overview}) enables effective accuracy-latency co-optimization on edge GPUs. By strategically constraining token budgets at these optimal lengths, we can maximize accuracy while minimizing latency penalties.
\begin{keytakeaway}
\textbf{Takeaway \#7:} Sequential scaling holds even when reasoning token control is applied. 
\end{keytakeaway}

% \begin{figure}[t]
%     \centering
%     % Subfigure 1
%     \begin{subfigure}[b]{0.48\linewidth}
%         \centering
%         \includegraphics[width=\linewidth]{new_accuracy/direct/direct_tokens.pdf}
%         \caption{Token usage}
%         \label{fig:direct_tokens}
%     \end{subfigure}
%     % \hspace{1em}
%     \hfill
%     % Subfigure 2
%     \begin{subfigure}[b]{0.48\linewidth}
%         \centering
%         \includegraphics[width=\linewidth]{new_accuracy/direct/direct_latency.pdf}
%         \caption{Latency}
%         \label{fig:direct_latency}
%     \end{subfigure}
%     % One caption for the whole figure
%     \caption{Accuracy versus average length and latency for direct \& reasoning models.}
%     \label{fig:direct_plot}
% \end{figure}

\textbf{Reasoning vs. non-reasoning models.} 
Fig.~\ref{fig:nr_latency} reveals distinct performance profiles between reasoning and non-reasoning approaches. The non-reasoning models direct generations (\symDirect{} markers: Qwen2.5-1.5B-it, Llama3.1-8B-it, Qwen2.5-14B-it) demonstrate competitive accuracy under low latency compared to the reasoning model counterparts. 
DS-Llama-8B's \emph{Base} configuration (\symBase{} markers \textcolor{DS-Llama-8B}{plum})  without token control achieves 5.7\% higher accuracy than the non-reasoning counterpart, \emph{Llama3.1-8B-it}, but at the cost of 13$\times$ longer runtime (87.2s vs 6.60s) as shown in Fig.~\ref{fig:latency_overview}. When \emph{DSR1-Llama-8B} is constrained to 128 tokens (\emph{128T}) to achieve sub-10s inference time, accuracy drops by 34\% compared to direct \emph{Llama3.1-8B-it}. Additionally, the direct \emph{Llama3.1-8B-it} consistently outperforms all 1.5B reasoning models configurations, establishing it as the preferred choice for latency budgets below 20 seconds.

 \begin{keytakeaway}
\textbf{Takeaway \#8:} Non-reasoning models offer a competitive latency-accuracy trade-off compared to reasoning models on a low token and latency budget.
\end{keytakeaway}

\subsection{Cost Analysis.}
% Figure~\ref{fig:acc_cost} show the accuracy cost tradeoff for selecting different inference strategies. It shows that, in general, better accuracy comes at a higher cost due to the use of larger reasoning models and longer output lengths.  It also provides guidance for model selection subject to price per token. When the budget is lower than \$0.01/1m tokens, ultra light-weight models like \emph{DSR1-Qwen-1.5B} and \emph{L1} are the only option. When the budget is between \$0.01/1m tokens and \$0.1/1m tokens non-reasoning models offer the best accuracy to cost tradeoff. Beyond a cost budget of \$0.1/1m, \emph{DSR1-Llama-8B} and \emph{DSR1-Qwen-14B} are both compelling options, but the users need to trade-off between model size and token length budget. 

Fig.~\ref{fig:cost_overview} illustrates the accuracy-cost trade-offs inherent in different inference strategies. The results confirm that superior accuracy typically incurs higher computational costs due to the deployment of larger reasoning models and extended output sequences. The analysis provides clear guidance for model selection based on token pricing constraints.
For budgets below \$0.01 per million tokens, ultra-lightweight models such as \emph{DSR1-Qwen-1.5B} and \emph{L1} represent the only viable options. Within the \$0.01-\$0.1 per million token range, non-reasoning models deliver optimal accuracy-to-cost ratios. Beyond \$0.1 per million tokens, both \emph{DSR1-Llama-8B} and \emph{DSR1-Qwen-14B} emerge as compelling alternatives, though users must carefully balance model size against token length budget constraints.

\begin{figure}[t]
    \centering
    %--- 128 ----------------------------------------------
    \begin{subfigure}[b]{0.48\linewidth}
        \centering
        \includegraphics[width=\linewidth]{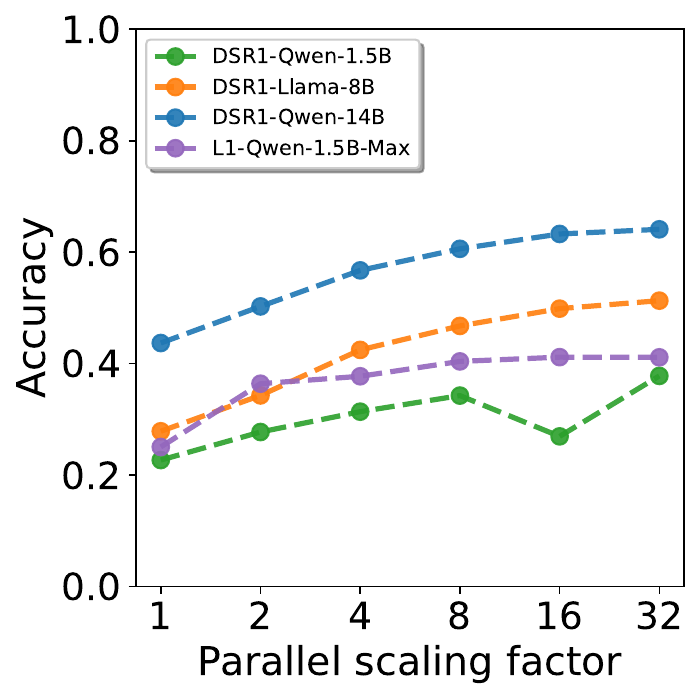}
            \caption{Output length = 128}
        \label{fig:ps_server_128}
    \end{subfigure}%
    \hfill
    %--- 512 ---------------------------------------------
    \begin{subfigure}[b]{0.48\linewidth}
        \centering
        \includegraphics[width=\linewidth]{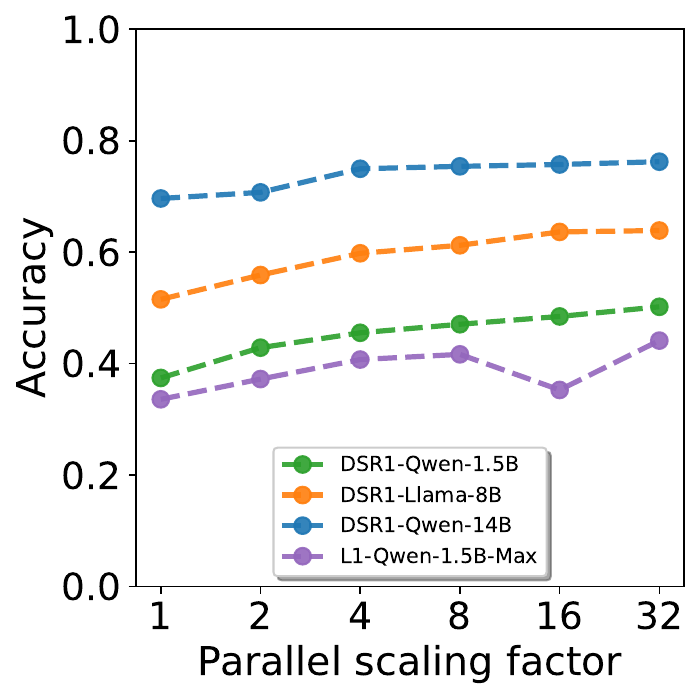}
        \caption{Output length = 512}
        \label{fig:ps_server_512}
    \end{subfigure}
    
    \caption{Accuracy vs.\ parallel scaling factor at output lengths 128 (a) and 512(b) on Full  MMLU-Redux.}
    \label{fig:ps_server_accuracy_comparison}
\end{figure}

\begin{figure}[t]
  \centering
  % Row 1: latency & energy
  \begin{subfigure}[b]{0.48\columnwidth}
    \centering
    \includegraphics[width=\linewidth]{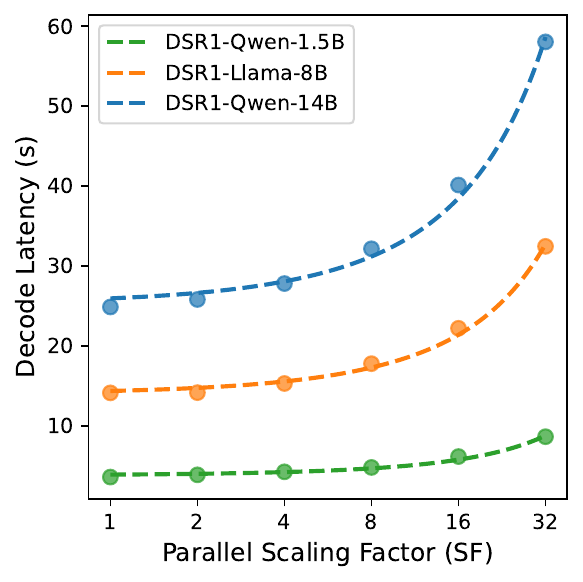}
    \caption{Decode latency}
    \label{fig:latency_decode_ps}
  \end{subfigure}%
  \hfill
  \begin{subfigure}[b]{0.48\columnwidth}
    \centering
    \includegraphics[width=\linewidth]{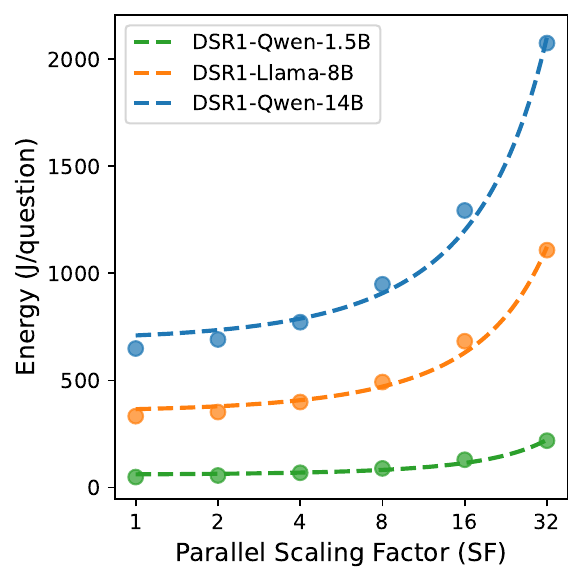}
    \caption{Energy per question}
    \label{fig:energy_decode_ps}
  \end{subfigure}

  \vspace{0.5em}

  % Row 2: power & memory utilization
  \begin{subfigure}[b]{0.57\columnwidth}
    \centering
    \includegraphics[width=\linewidth]{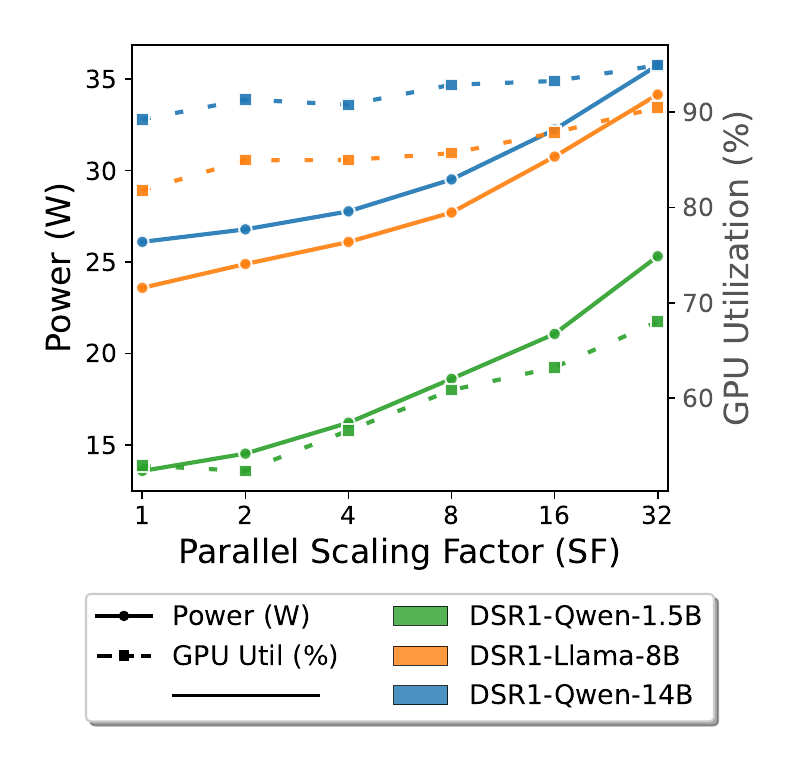}
    \caption{Power and GPU utilization}
    \label{fig:power_gpu_decode_ps}
  \end{subfigure}

  \caption{Parallel-scaling on Orin: 
    (a) decode latency, 
    (b) energy per question, 
    (c) power and GPU utilization}
  \label{fig:ps_orin_combined}
\end{figure}

\subsection{Parallel Test Time Scaling}
\label{sec:scaling}
The preceding studies examined single-batch inference without parallel scaling. As discussed in Section~\ref{sec:background}, parallel scaling represents another test-time scaling approach that can increase accuracy with minimal latency overhead. 
We now quantify how parallel scaling impacts latency, power, and energy efficiency across our target models.
For consistency, all experiments use a fixed 128 output token budget.  
The prefill phase is executed once with a batch size of~1; during the decode phase we increase the batch size to match the target parallelization factor.  
Results from the parallel decoders are combined with a lightweight majority-voting scheme to produce the final answer.

\textbf{Impact on Accuracy.} 
% A full  MMLU-Redux benchmark study with generation batching using  server class GPUs to create a baseline accuracy comparison.
% Figure \ref{fig:ps_server_128} shows that when increasing the scale factor from 1× to 32×, accuracy rises by roughly 1.5 × to 1.8 × across both sizes of models with 128 output token budget. However, in Figure \ref{fig:ps_server_512}, with the output budget raised to 512 tokens, accuracy gains plateau after just 4× scaling for larger models and smaller increases for smaller models, indicating that at this larger budget, sequential scaling drives accuracy and additional parallel samples yield diminishing returns. We also observe that models fined-tuned for length control such as L1-Qwen-1.5B-Max models finetuned do not benefit from increased parallel scaling beyond 2x at a 128 output token budget and 8x at a 512 output token budget. Smaller models exhibit a drop in accuracy at 16x scaling factor.
First, we study how parallel scaling impacts the accuracy by evaluating on MMLU-Redux.
Fig.\ref{fig:ps_server_128} shows that scaling from 1× to 32× yields accuracy improvements of approximately 1.5× to 1.8× across both model sizes under a 128-token output budget. Conversely, Fig.\ref{fig:ps_server_512} reveals different behavior when the output budget increases to 512 tokens: accuracy gains plateau after only 4× scaling for larger models, with even more limited improvements observed in smaller models. This plateau effect indicates that under higher token budgets, sequential scaling becomes the dominant factor driving accuracy improvements, while additional parallel samples produce diminishing returns.
Models fine-tuned for length control exhibit distinct behavior. The L1-Qwen-1.5B-Max variants show negligible benefits from parallel scaling beyond 2× (128-token budget) and 8× (512-token budget). Furthermore, smaller models experience accuracy degradation at the 16× scaling factor.

% , which we attribute to a transition noise.

\textbf{Impact on Decode Latency.}  
% While decoding complexity scales linearly with batch size in theory, its computational and memory components exhibit distinct scaling rates. At low batch sizes (memory-bound regimes), increasing SF amplifies memory traffic and thus theoretically increases latency. However, Tensor Core on GPUs introduces a potential optimization: the initial scaling steps may incur minimal overhead due to batch dimension padding in 128-size blocks.
Fig.~\ref{fig:latency_decode_ps} presents an ablation study of decoding latency versus parallel scaling factors (SF) on the NVIDIA Jetson Orin platform.
Since both the compute and memory complexity of decoding scales with batch size, in theory larger batch size should lead to higher decode latency. However, Tensor Core on GPUs introduces a potential optimization: the initial scaling steps may incur minimal overhead due to batch dimension padding in 128-size blocks.
Fig.~\ref{fig:latency_decode_ps} reveals a slight latency increase for SF $<$ 128, with latency rising approximately 2× from SF=1 to SF=64 across all models. While this modest increase partially validates the Tensor Core hypothesis, the non-flat latency profile demonstrates that scaling isn't completely free. 
\textbf{Impact on Power and Energy.} 
Fig.\ref{fig:power_gpu_decode_ps} illustrates how average GPU power consumption varies with parallel scaling factors and overall GPU utilization. Power consumption increases substantially with parallel scaling, rising from 14W to 25W for the 1.5B model and from approximately 25W to 35W for the larger 8B and 14B models. These discrete power trends correspond to distinct GPU power states triggered by different utilization levels, as shown on the secondary axis. This scaling behavior aligns with the increased computational and memory complexity introduced by parallel batching.

\begin{figure}[t]
    \centering
    \begin{subfigure}[b]{0.45\linewidth}
        \centering
        \includegraphics[width=\linewidth]{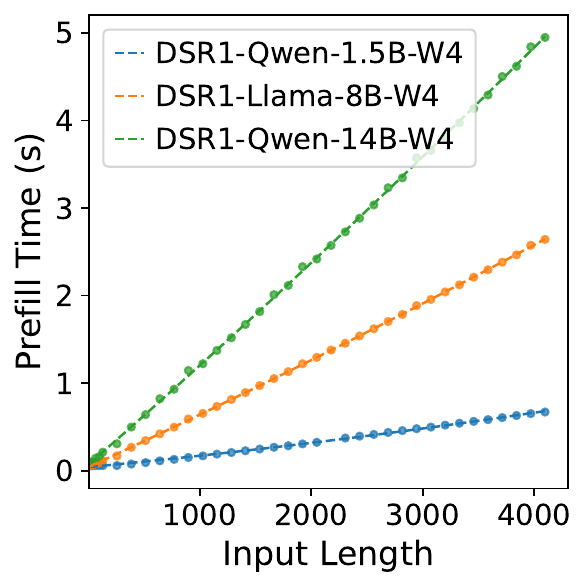}
        \caption{Prefill Time}
        \label{fig:prefill_latency_quant}
    \end{subfigure}
    \hfill
    \begin{subfigure}[b]{0.45\linewidth}
        \centering
        \includegraphics[width=\linewidth]{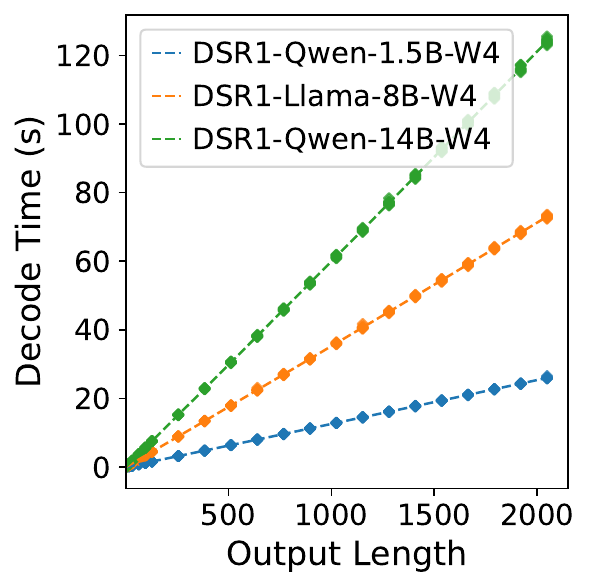}
        \caption{Decode Time}
        \label{fig:decode_lantency_quant}
    \end{subfigure}
    \caption{Prefill (left) and decode phase latency (right) as a function of sequence length for the quantized models.}
    \label{fig:prefill_decode_lantency_quant}
\end{figure}

Fig.\ref{fig:energy_decode_ps} demonstrates how energy per question varies with parallel scaling factor across the three models. The energy consumption follows a similar trend to decode latency, as longer inference times naturally result in higher energy consumption, particularly when power draw is simultaneously increasing. For the 14B model, energy per question increases modestly by less than 1.5× from SF=1 to SF=4, indicating efficient resource utilization in this range. However, at SF=16, energy consumption doubles, reflecting the transition to higher energy overhead with parallel scaling. 
% hese findings suggest that parallel scaling offers an attractive energy-accuracy trade-off at moderate scaling factors, but efficiency gains could diminish beyond the optimal operating range.

\begin{keytakeaway}
\textbf{Takeaway \#9}: Parallel scaling improves accuracy with minimal latency and energy overhead  at small scaling factors ($\le$ 8).  
\end{keytakeaway} 
\vspace{1.5em}
\textbf{Impact on Utilization.} 
% % Figure \ref{fig:power_gpu_decode_ps} shows that aggregate GPU utilization rises almost perfectly linearly with the parallel scale factor---from $1\times$ all the way to $32\times$---mirroring increased tensor-core utilization. DRAM bandwidth utilization, by contrast, remains essentially flat. CPU utilization holds steady $\leq$ 20\% regardless of scale factor, revealing a large pool of idle host-side compute. In other words, generation batching trades extra latency and energy for proportional gains in on-chip compute efficiency, and further latency reductions can be unlocked by offloading lightweight graph kernels---tokenization, layer-norm, softmax, embedding lookups---to the host CPU and overlapping them with GPU matmuls. Due to the shared memory nature of Orin's SoC this would present minimal communication overheads. Moreover, on Jetson Orin the dedicated deep-learning (DLA) and programmable vision (PVA) accelerators sit unused during transformer inference; exploring how to map parts of the attention/FFN workload onto these engines could yield additional throughput and energy-efficiency wins.
Fig.\ref{fig:power_gpu_decode_ps} shows that GPU utilization rises linearly with the parallel scale factor. 
% DRAM bandwidth utilization, by contrast, remains essentially flat.
DRAM read bandwidth dominates---rising above $\sim$20\% on the 1.5B model and above $\sim$60\% on the 14B model---since decode kernels continually fetch weights, and activation tiles from DRAM. Write bandwidth stays below 10\%, reflecting KV-cache write back and output logits commits. 

CPU utilization holds steady $\leq$ 20\% regardless of scale factor, revealing a large pool of idle host-side compute. In other words, generation batching trades extra latency and energy for proportional gains in on-chip compute efficiency, and further latency reductions can be unlocked by offloading lightweight graph kernels---tokenization, layer-norm, softmax, embedding lookups---to the host CPU and overlapping them with GPU matmuls. Due to the shared memory nature of Orin's SoC this would present minimal communication overheads. Moreover, on Jetson Orin the dedicated deep-learning (DLA) and programmable vision (PVA) accelerators sit unused during transformer inference; exploring how to map parts of the attention/FFN workload onto these engines could yield additional throughput and energy-efficiency wins.

% \textbf{Takeaway \#:} {Transformer inference on SoC architectures leaves a gap in full compute utilization. Parallel scaling is one method that narrows the gap with optimal points for energy and latency}

\begin{keytakeaway}
\textbf{Takeaway \#10:} {Parallel scaling utilizes hardware resources effectively and improves the overall GPU utilization.}
\end{keytakeaway}

% \input{3_4_PS_w_L1}
% File: 3_3_quantization.tex

  \begin{figure}[t]
    \centering
    \begin{subfigure}[b]{0.48\linewidth}
        \centering
        \includegraphics[width=\linewidth]{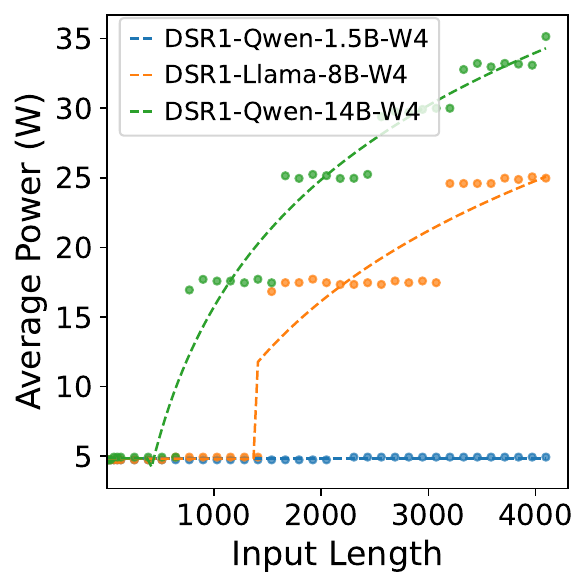}
        \caption{Power}
        \label{fig:prefill_power_quant}
    \end{subfigure}
    \hfill
    \begin{subfigure}[b]{0.48\linewidth}
        \centering
        \includegraphics[width=\linewidth]{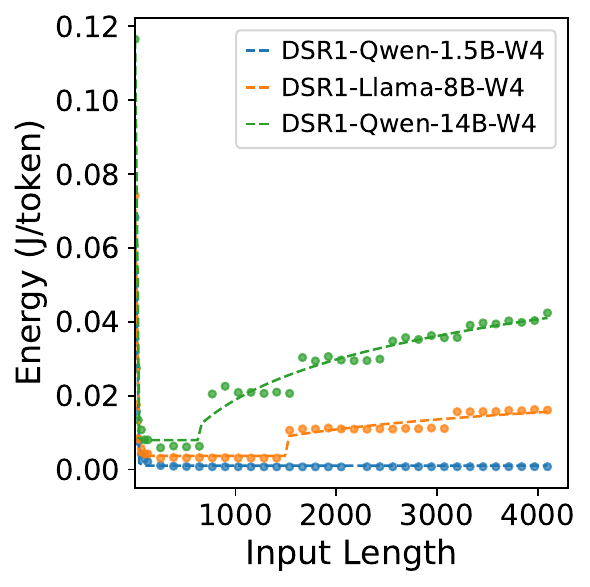}
        \caption{Energy/token}
        \label{fig:prefill_energy_per_token}
    \end{subfigure}
    \caption{Prefill phase power (left) and energy/token (right) as a function of sequence length for the quantized models.}
    \label{fig:prefill_energy_power_quant}
\end{figure}

\subsection{Impact of Quantization}

% : DeepSeek-R1-Distill-Qwen-14B, DeepSeek-R1-Distill-Qwen-1.5B and DeepSeek-R1-Distill-Llama-8B 
% across three experimental scenarios:

% \begin{itemize}
%     \item \textbf{Prefill:} Latency and energy consumption are measured for input lengths of 1, 8, 32, 64, and multiples of 128 up to 4096 during the prompt processing phase.
%     \item \textbf{Decode:} For input lengths that are multiples of 128 up to 1024, we sweep output lengths (1, 16, 64, 96, and multiples of 128 up to 2045) and measure latency and energy per token during the generation phase.
%     \item \textbf{MMLU-Redux Performance:} Accuracy impact is assessed using the MMLU-Redux benchmark to quantify any loss due to quantization.
% \end{itemize}

% Fig.~\ref{fig:prefill_decode_lantency_quant} shows the prefill and decode latency for the quantized small reasoning models. 
% Fig.~\ref{fig:prefill_energy_power_quant} illustrates prefill phase power and energy per token across input lengths.

% (Detailed numbers are also listed in Appendix~\ref{} Table~\ref{tab:prefill_base_vs_awq}). 
% Table~\ref{tab:prefill_base_vs_awq} summarizes base and quantized model performance for prefill.
% Fig.~\ref{fig:decode_energy_power_combined} shows decode phase power and energy per token.
% Table~\ref{tab:filtered512} presents decode performance at 512 input tokens.
% The decode power profile follows the same behavior described in Eqn. \ref{eqn:power_decode}.

% Both accuracy and latency results from 
% Fig. \ref{fig:quantization_performance} shows that smaller size model does not benefit from quantization compared to the bigger models. 

We evaluate the effect of quantization on reasoning models by applying W4A16 (4-bit weights, 16-bit activations) using the LLM Compressor AWQ configuration in vLLM. On the Jetson Orin GPU, however, computation falls back to INT8 since its Ampere architecture does not support INT4.

\begin{figure}[t]
    \centering
    \begin{subfigure}[b]{0.42\linewidth}
        \centering
        \includegraphics[width=\linewidth]{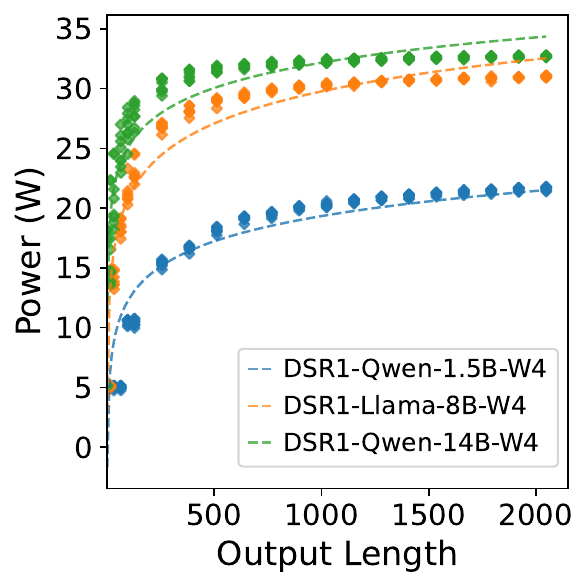}
        \caption{Power}
        \label{fig:decode_power_quantized}
    \end{subfigure}
    \hfill
    \begin{subfigure}[b]{0.42\linewidth}
        \centering
        \includegraphics[width=\linewidth]{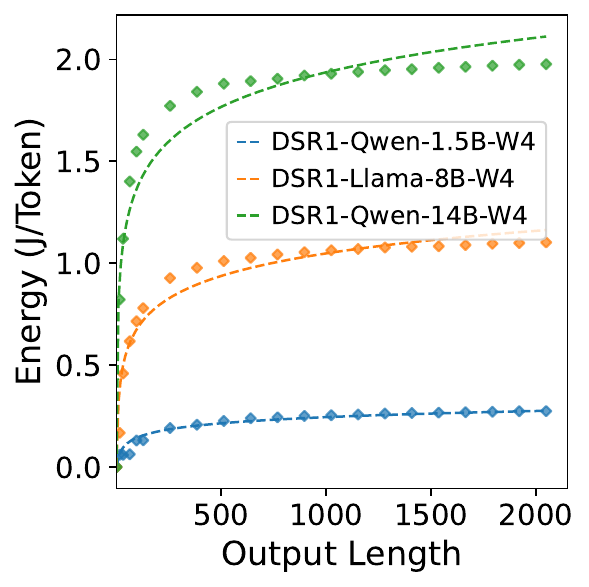}
        \caption{Energy/token}
        \label{fig:decode_energy_per_token}
    \end{subfigure}
    \caption{Decode phase power (left) and energy/token (right) as a function of sequence length at 512 input length.}
    \label{fig:decode_energy_power_combined}
    % \vspace{-0.4cm}
\end{figure}

Fig.~\ref{fig:prefill_decode_lantency_quant} presents prefill and decode latency for the quantized models, while Fig.~\ref{fig:prefill_energy_power_quant} and~\ref{fig:decode_energy_power_combined} show power and energy per token during prefill and decode. They have a shorter prefill and decode time at lower energy/token compared to their non-quantized models shown in Fig. \ref{fig:prefill-time} and \ref{fig:decode_latency}.

\begin{figure}[t]
    \centering

    % --- Subfigure 1: Accuracy ---
    \begin{subfigure}{0.42\linewidth}
        \includegraphics[width=\textwidth]{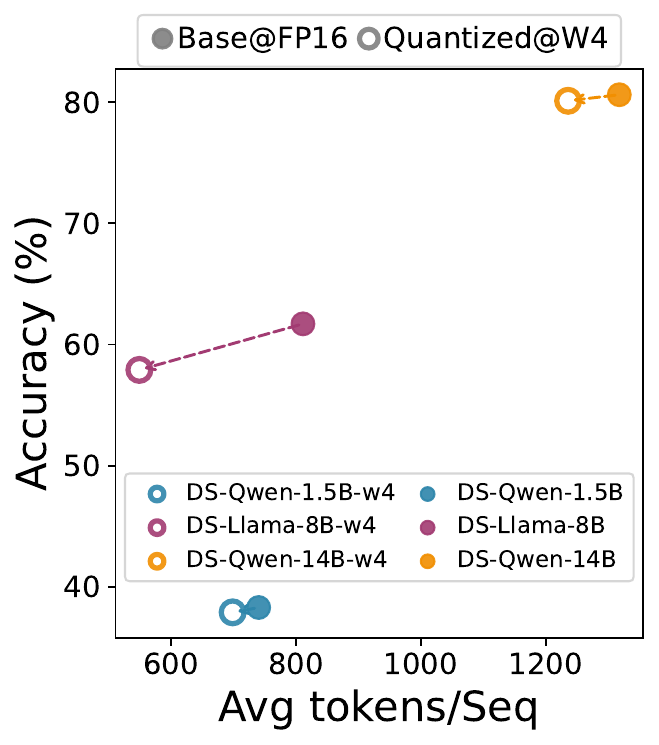}
        \caption{Average Ouput Tokens}
        \label{fig:accuracy_quantized}
    \end{subfigure}
    \hfill 
    % --- Subfigure 2: Latency ---
    \begin{subfigure}{0.42\linewidth}
        \includegraphics[width=\textwidth]{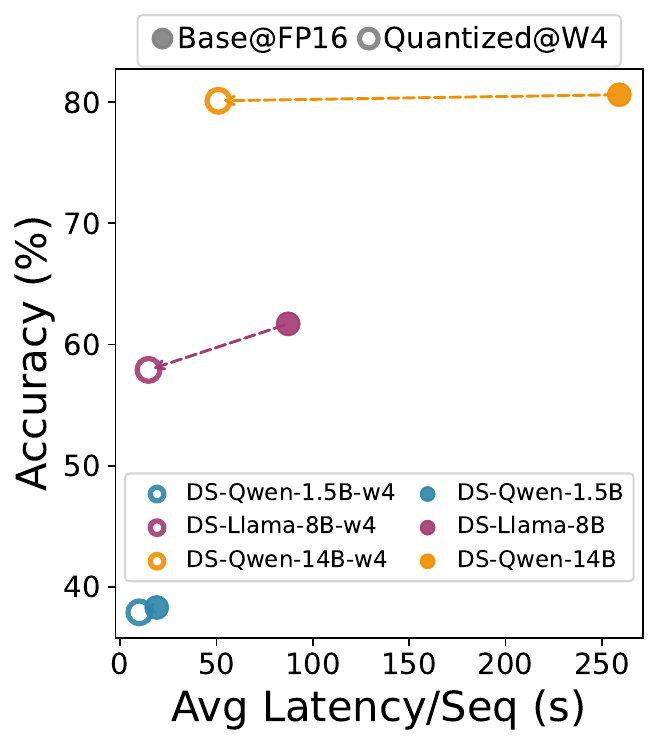}
        \caption{Average Latency(s)}
        \label{fig:latency_quantized}
    \end{subfigure}

    \caption{Comparison of quantized and non-quantized models on accuracy, average output token length, and latency.}
    \label{fig:quantization_performance}
    % \vspace{-0.5cm}
\end{figure}

Fig.~\ref{fig:quantization_performance} demonstrates that AWQ quantization reduces accuracy relative to FP16 by 
(i) DSR1-Qwen-1.5B: -1.04\%, (ii) DSR1-Llama-8B: -6.16\% and DSR1-Qwen-14B: -0.62\% of relative loss.
Finally, Figure~\ref{fig:accuracy_quantized} indicates that quantized models generate fewer decoding tokens than their FP16 counterparts. As a result, quantization improves latency by roughly 2–5×, with larger models benefiting more than smaller ones, as shown in Fig.\ref{fig:quantization_performance}.

\begin{keytakeaway}
{\bfseries Takeaway \#11:}
AWQ-based W4 quantization improves latency and reduces energy per token with minor accuracy loss. Gains increase with larger model size. 
\end{keytakeaway}

\subsection{Impact of Inference Frameworks}
In this section, we present a latency comparison across popular inference frameworks, including the Hugging Face Transformers library (HFT)~\cite{wolf2019huggingface}, vLLM~\cite{kwon2023efficient} and TRT-LLM. We evaluate end-to-end inference time using three input–output sequence length combinations on DSR1-Llama-8B model, and observe that vLLM(v0.86) achieves a speedup of 1.11$\times$ to 1.13$\times$ over HFT(v4.46.2) and a similar performance when compared to TRT-LLM (v0.12).

% \todo{Add the version of vllm and huggingface transformers.}
% \todo{Low Priority: add comparison to SGLang https://github.com/shahizat/SGLang-Jetson}
% \todo{Low Priority: make it comprehensive and add 1.5B and 14B models}

\begin{table}[t]
    \caption{Inference Engine Performance Comparison on  DeepSeek-R1-Distill-Llama-8B}
    \label{tab:inference_engines}
    \centering
    \resizebox{1\columnwidth}{!} {

\begin{tabular}{|c|c|c|}
    \hline
    \multirow{2}{*}{\textbf{Input Length}} & \multirow{2}{*}{\textbf{Output Length}} & \textbf{Latency (s)} \\ % Changed ms to s for consistency
    & & \textbf{HF $\to$ vLLM $\to$ TRT-LLM (Speedup)} \\ % Clarified Speedup
    \hline
    16 & 128 & \num{14.23} $\to$ \num{12.73} (\num{1.12}\texttimes) $\to$ \num{12.79} (\num{1.00}\texttimes) \\ % Fixed 12794 and parentheses
    \hline % Added \hline for better table structure
    64 & 128 & \num{14.29} $\to$ \num{12.75} (\num{1.12}\texttimes) $\to$ \num{12.46} (\num{1.05}\texttimes) \\ % Fixed 12459
    \hline % Added \hline
    128 & 128 & \num{14.41} $\to$ \num{12.78} (\num{1.13}\texttimes) $\to$ \num{12.88} (\num{0.99}\texttimes) \\ % Fixed 12876
    \hline
\end{tabular}}
\end{table}

\section{Discussion and Future Work}

% \subsection{GPU Optimizations}

Our study reveals significant opportunities for co-optimizing GPU architecture and software to enhance edge inference performance for reasoning LLMs. 
% Despite implementing parallel scaling with a factor of 64, peak GPU utilization remains at approximately 50\%, indicating substantial untapped computational capacity.
The bandwidth-bound nature of reasoning LLM inference becomes evident when examining the operational characteristics of the Jetson AGX Orin platform. With a FLOPs-to-bytes ratio of approximately 1375 for fp16 tensor operations—significantly higher than the operational intensity of batch size 1 GEMV operations—the system is constrained by memory bandwidth rather than computational throughput. This bottleneck is particularly pronounced in reasoning LLMs where decoding operations dominate 99\% of the inference time, creating a critical need for enhanced memory bandwidth to achieve optimal performance.
Beyond GPU utilization, our analysis reveals that other computational resources within the Orin SoC remain underutilized during inference. Both ARM CPU cores and DLA units present opportunities for performance optimization through heterogeneous computing approaches.
% workload distribution and

Several optimization strategies warrant investigation to address these performance limitations. \textbf{Quantization}~\cite{xiao2023smoothquant, yao2023zeroquant, lin2023awq} can reduce model precision to 4-bit or lower while maintaining accuracy. \textbf{Kernel fusion} \cite{dao2022flashattention,li2022automatic, zheng2020ansor} can minimize memory traffic by combining not only attention operations but also normalization, activation functions, and other tensor operations into unified kernels. \textbf{Prefetching}~\cite{yuzuguler2025preserve} can overlap memory transfers with computation to hide latency. \textbf{Speculative decoding} \cite{leviathan2023fast, chen2023accelerating, li2024eagle} can increase computational intensity by predicting multiple tokens in parallel. 
% Finally, heterogeneous computing approaches can leverage ARM CPU and DLA cores for complementary tasks such as sampling, beam search, and post-processing operations. 
These optimizations, combined with inference-time scaling strategies, offer a comprehensive approach to maximizing reasoning LLM performance on edge devices.

\section{Conclusion}
This work presents a comprehensive characterization of Large Language Model (LLM) reasoning workloads on edge GPU platforms. We systematically quantify the impact of model scale, input/output sequence lengths, and inference-time scaling techniques on latency, power consumption, and energy efficiency. By deriving analytical models that map these parameters to performance metrics, we enable rapid evaluation of optimal deployment strategies without exhaustive hardware testing. Furthermore, we assess token control methodologies for multi-step reasoning tasks, characterizing their fundamental latency-accuracy tradeoffs. Our analysis demonstrates the superior cost-effectiveness of edge deployment for LLM reasoning and provides concrete configuration guidelines for maximizing accuracy under diverse latency constraints. These findings deliver both practical deployment frameworks and fundamental insights for efficient edge AI systems.
\clearpage
\bibliography{refs}

\begin{thebibliography}{10}

\bibitem{aggarwal2025l1}
Pranjal Aggarwal and Sean Welleck.
\newblock L1: Controlling how long a reasoning model thinks with reinforcement learning.
\newblock {\em arXiv preprint arXiv:2503.04697}, 2025.

\bibitem{brown2024large}
Bradley Brown, Jordan Juravsky, Ryan Ehrlich, Ronald Clark, Quoc~V Le, Christopher R{\'e}, and Azalia Mirhoseini.
\newblock Large language monkeys: Scaling inference compute with repeated sampling.
\newblock {\em arXiv preprint arXiv:2407.21787}, 2024.

\bibitem{butcher2025precise}
Bradley Butcher, Michael O’Keefe, and James Titchener.
\newblock Precise length control for large language models.
\newblock {\em Natural Language Processing Journal}, 11:100143, 2025.

\bibitem{chen2023accelerating}
Charlie Chen, Sebastian Borgeaud, Geoffrey Irving, Jean-Baptiste Lespiau, Laurent Sifre, and John Jumper.
\newblock Accelerating large language model decoding with speculative sampling.
\newblock {\em arXiv preprint arXiv:2302.01318}, 2023.

\bibitem{dao2022flashattention}
Tri Dao, Daniel~Y Fu, Stefano Ermon, Atri Rudra, and Christopher R{\'e}.
\newblock Flashattention: Fast and memory-efficient exact attention with io-awareness.
\newblock {\em Advances in Neural Information Processing Systems}, 35:16344--16359, 2022.

\bibitem{deepseek_r1_0528}
{DeepSeek}.
\newblock Deepseek-r1 (0528) via openrouter.
\newblock \url{https://openrouter.ai/deepseek/deepseek-r1-0528}, 2024.
\newblock Accessed: 2025-06-23.

\bibitem{guo2025deepseek}
DeepSeek-AI.
\newblock Deepseek-r1: Incentivizing reasoning capability in llms via reinforcement learning.
\newblock {\em arXiv preprint arXiv:2501.12948}, 2025.

\bibitem{huggingface2024scaling}
{Edward Beeching, Lewis Tunstall, Sasha Rush}.
\newblock Scaling test-time compute efficiently.
\newblock \url{https://huggingface.co/spaces/HuggingFaceH4/blogpost-scaling-test-time-compute}, 2024.
\newblock Accessed: 2025-06-24.

\bibitem{gao2024interpretable}
Zitian Gao, Boye Niu, Xuzheng He, Haotian Xu, Hongzhang Liu, Aiwei Liu, Xuming Hu, and Lijie Wen.
\newblock Interpretable contrastive monte carlo tree search reasoning.
\newblock {\em arXiv preprint arXiv:2410.01707}, 2024.

\bibitem{gema2024mmlu}
Aryo~Pradipta Gema, Joshua Ong Jun~Leang, Giwon Hong, Alessio Devoto, Alberto Carlo~Maria Mancino, Rohit Saxena, Xuanli He, Yu~Zhao, Xiaotang Du, Mohammad~Reza Ghasemi~Madani, Claire Barale, Robert McHardy, Joshua Harris, Jean Kaddour, Emile van Krieken, and Pasquale Minervini.
\newblock Are we done with mmlu?
\newblock In {\em arXiv preprint arXiv:2406.04127}, 2024.

\bibitem{grattafiori2024llama}
Meta GenAI.
\newblock The llama 3 herd of models.
\newblock {\em arXiv preprint arXiv:2407.21783}, 2024.

\bibitem{han2024token}
Tingxu Han, Zhenting Wang, Chunrong Fang, Shiyu Zhao, Shiqing Ma, and Zhenyu Chen.
\newblock Token-budget-aware llm reasoning.
\newblock {\em arXiv preprint arXiv:2412.18547}, 2024.

\bibitem{hendrycks2021mmlu}
Dan Hendrycks, Collin Burns, Steven Basart, Andy Zou, Mantas Mazeika, Dawn Song, and Jacob Steinhardt.
\newblock Measuring massive multitask language understanding.
\newblock In {\em International Conference on Learning Representations (ICLR)}, 2021.

\bibitem{hooper2025ets}
Coleman Hooper, Sehoon Kim, Suhong Moon, Kerem Dilmen, Monishwaran Maheswaran, Nicholas Lee, Michael~W Mahoney, Sophia Shao, Kurt Keutzer, and Amir Gholami.
\newblock Ets: Efficient tree search for inference-time scaling.
\newblock {\em arXiv preprint arXiv:2502.13575}, 2025.

\bibitem{kwon2023efficient}
Woosuk Kwon, Zhuohan Li, Siyuan Zhuang, Ying Sheng, Lianmin Zheng, Cody~Hao Yu, Joseph~E. Gonzalez, Hao Zhang, and Ion Stoica.
\newblock Efficient memory management for large language model serving with pagedattention.
\newblock In {\em Proceedings of the ACM SIGOPS 29th Symposium on Operating Systems Principles}, 2023.

\bibitem{leviathan2023fast}
Yaniv Leviathan, Matan Kalman, and Yossi Matias.
\newblock Fast inference from transformers via speculative decoding.
\newblock {\em arXiv preprint arXiv:2211.17192}, 2023.

\bibitem{li2022automatic}
Ao~Li, Bojian Zheng, Gennady Pekhimenko, and Fan Long.
\newblock Automatic horizontal fusion for gpu kernels.
\newblock In {\em 2022 IEEE/ACM International Symposium on Code Generation and Optimization (CGO)}, pages 14--27. IEEE, 2022.

\bibitem{li2024eagle}
Yuhui Li, Fangyun Wei, Chao Zhang, and Hongyang Zhang.
\newblock Eagle: Speculative sampling requires rethinking feature uncertainty.
\newblock {\em arXiv preprint arXiv:2401.15077}, 2024.

\bibitem{lin2023awq}
Ji~Lin, Jiaming Tang, Haotian Tang, Shang Yang, Xingyu Dang, and Song Han.
\newblock Awq: Activation-aware weight quantization for llm compression and acceleration.
\newblock {\em arXiv preprint arXiv:2306.00978}, 2023.

\bibitem{deepcoder2025}
Michael Luo, Sijun Tan, Roy Huang, Ameen Patel, Alpay Ariyak, Qingyang Wu, Xiaoxiang Shi, Rachel Xin, Colin Cai, Maurice Weber, Ce~Zhang, Li~Erran Li, Raluca~Ada Popa, and Ion Stoica.
\newblock Deepcoder: A fully open-source 14b coder at o3-mini level.
\newblock https://pretty-radio-b75.notion.site/DeepCoder-A-Fully-Open-Source-14B-Coder-at-O3-mini-Level-1cf81902c14680b3bee5eb349a512a51, 2025.

\bibitem{luo2025deepscaler}
Michael Luo, Sijun Tan, Justin Wong, Xiaoxiang Shi, William~Y Tang, Manan Roongta, Colin Cai, Jeffrey Luo, Tianjun Zhang, Li~Erran Li, et~al.
\newblock Deepscaler: Surpassing o1-preview with a 1.5 b model by scaling rl.
\newblock 2025.

\bibitem{ma2025reasoning}
Wenjie Ma, Jingxuan He, Charlie Snell, Tyler Griggs, Sewon Min, and Matei Zaharia.
\newblock Reasoning models can be effective without thinking.
\newblock {\em arXiv preprint arXiv:2504.09858}, 2025.

\bibitem{abdin2024phi}
Microsoft.
\newblock Phi-3 technical report: A highly capable language model locally on your phone.
\newblock {\em arXiv preprint arXiv:2404.14219}, 2024.

\bibitem{muennighoff2025s1}
Niklas Muennighoff, Zitong Yang, Weijia Shi, Xiang~Lisa Li, Li~Fei-Fei, Hannaneh Hajishirzi, Luke Zettlemoyer, Percy Liang, Emmanuel Cand{\`e}s, and Tatsunori Hashimoto.
\newblock s1: Simple test-time scaling.
\newblock {\em arXiv preprint arXiv:2501.19393}, 2025.

\bibitem{nvidiaJetsonOrin}
{NVIDIA Corporation}.
\newblock {Jetson AGX Orin} series – technical specifications.
\newblock https://developer.nvidia.com/embedded/jetson-agx-orin, 2022.
\newblock Available online at NVIDIA Developer (Accessed June 2025).

\bibitem{openai_o1_preview}
{OpenAI}.
\newblock Openai o1-preview (2024-09-12) via openrouter.
\newblock https://openrouter.ai/openai/o1-preview-2024-09-12, 2024.

\bibitem{jaech2024openai}
OpenAI.
\newblock Openai o1 system card.
\newblock {\em arXiv preprint arXiv:2412.16720}, 2024.

\bibitem{openai_api_pricing}
{OpenAI}.
\newblock Openai api pricing.
\newblock \url{https://platform.openai.com/docs/pricing}, 2025.

\bibitem{snell2024scaling}
Charlie Snell, Jaehoon Lee, Kelvin Xu, and Aviral Kumar.
\newblock Scaling llm test-time compute optimally can be more effective than scaling model parameters.
\newblock {\em arXiv preprint arXiv:2408.03314}, 2024.

\bibitem{sui2503stop}
Yang Sui, Yu-Neng Chuang, Guanchu Wang, Jiamu Zhang, Tianyi Zhang, Jiayi Yuan, Hongyi Liu, Andrew Wen, Shaochen Zhong, Hanjie Chen, et~al.
\newblock Stop overthinking: A survey on efficient reasoning for large language models, 2025.
\newblock {\em URL https://arxiv. org/abs/2503.16419}.

\bibitem{team2023gemini}
Gemini Team.
\newblock Gemini: a family of highly capable multimodal models.
\newblock {\em arXiv preprint arXiv:2312.11805}, 2023.

\bibitem{team2024gemma}
Gemma~Google Team.
\newblock Gemma: Open models based on gemini research and technology.
\newblock {\em arXiv preprint arXiv:2403.08295}, 2024.

\bibitem{qwen2.5}
Qwen Team.
\newblock Qwen2.5: A party of foundation models, September 2024.

\bibitem{yang2025qwen3}
Qwen Team.
\newblock Qwen3 technical report.
\newblock {\em arXiv preprint arXiv:2505.09388}, 2025.

\bibitem{wang2025faster}
Zili Wang, Tianyu Zhang, Haoli Bai, Lu~Hou, Xianzhi Yu, Wulong Liu, Shiming Xiang, and Lei Zhu.
\newblock Faster and better llms via latency-aware test-time scaling.
\newblock {\em arXiv preprint arXiv:2505.19634}, 2025.

\bibitem{wei2022chain}
Jason Wei, Xuezhi Wang, Dale Schuurmans, Maarten Bosma, Brian Ichter, Fei Xia, Ed~H. Chi, Quoc~V. Le, and Denny Zhou.
\newblock Chain-of-thought prompting elicits reasoning in large language models.
\newblock In {\em Advances in Neural Information Processing Systems (NeurIPS)}, 2022.

\bibitem{wolf2019huggingface}
Thomas Wolf, Lysandre Debut, Victor Sanh, Julien Chaumond, Clement Delangue, Anthony Moi, Pierric Cistac, Tim Rault, R{\'e}mi Louf, Morgan Funtowicz, et~al.
\newblock Huggingface's transformers: State-of-the-art natural language processing.
\newblock {\em arXiv preprint arXiv:1910.03771}, 2019.

\bibitem{wu2024inference}
Yangzhen Wu, Zhiqing Sun, Shanda Li, Sean Welleck, and Yiming Yang.
\newblock Inference scaling laws: An empirical analysis of compute-optimal inference for problem-solving with language models.
\newblock {\em arXiv preprint arXiv:2408.00724}, 2024.

\bibitem{xiao2023smoothquant}
Guangxuan Xiao, Ji~Lin, Mickael Seznec, Hao Wu, Julien Demouth, and Song Han.
\newblock Smoothquant: Accurate and efficient post-training quantization for large language models.
\newblock {\em International Conference on Machine Learning}, pages 38087--38099, 2023.

\bibitem{yao2023zeroquant}
Zhewei Yao, Reza Yazdani~Aminabadi, Minjia Zhang, Xiaoxia Wu, Conglong Li, and Yuxiong He.
\newblock Zeroquant-v2: Exploring post-training quantization in llms from comprehensive study to low rank compensation.
\newblock {\em arXiv preprint arXiv:2303.08302}, 2023.

\bibitem{yuzuguler2025preserve}
Ahmet~Caner Y{\"u}z{\"u}g{\"u}ler, Jiawei Zhuang, and Lukas Cavigelli.
\newblock Preserve: Prefetching model weights and kv-cache in distributed llm serving.
\newblock {\em arXiv preprint arXiv:2501.08192}, 2025.

\bibitem{zheng2024natural}
Huaixiu~Steven Zheng, Swaroop Mishra, Hugh Zhang, Xinyun Chen, Minmin Chen, Azade Nova, Le~Hou, Heng-Tze Cheng, Quoc~V Le, Ed~H Chi, et~al.
\newblock Natural plan: Benchmarking llms on natural language planning.
\newblock {\em arXiv preprint arXiv:2406.04520}, 2024.

\bibitem{zheng2020ansor}
Lianmin Zheng, Chengfan Jia, Minmin Sun, Zhao Wu, Cody~Hao Yu, Ameer Haj-Ali, Yida Wang, Jun Yang, Danyang Zhuo, Koushik Sen, et~al.
\newblock Ansor: Generating high-performance tensor programs for deep learning.
\newblock {\em 14th USENIX Symposium on Operating Systems Design and Implementation}, pages 863--879, 2020.

\bibitem{zhou2019edge}
Zhi Zhou, Xu~Chen, En~Li, Liekang Zeng, Ke~Luo, and Junshan Zhang.
\newblock Edge intelligence: Paving the last mile of artificial intelligence with edge computing.
\newblock {\em Proceedings of the IEEE}, 107(8):1738--1762, 2019.

\end{thebibliography}
\clearpage
\appendices
% File: 7_appendix.tex

 \lstdefinestyle{codebox}{%
  frame=single,
  rulecolor=\color{black!50},
  framerule=0.6pt,
  backgroundcolor=\color{black!3},
  basicstyle=\ttfamily\small,
  numbers=none,
  xleftmargin=1em,
  xrightmargin=1em,
  framexleftmargin=1em,
  breaklines=false,      
  breakatwhitespace=false,
  columns=fullflexible,
  keepspaces=true,
  showstringspaces=false,
  tabsize=2
}
\lstset{style=codebox}

\section{Artifact Appendix: EdgeReasoning}
\subsection{Abstract}
This section describes how to obtain the EdgeReasoning artifact and reproduce the key results in the paper. The artifact is validated on NVIDIA Jetson Orin AGX (ARM64 + JetPack 6.2/CUDA 12.8) and \texttt{x86\textunderscore64} servers with NVIDIA GPUs; other systems may work but were not evaluated.

\subsection{Artifact check-list (meta-information)}
\begin{itemize}
    \item \textbf{Algorithm:}  Evaluation of LLM inference on edge systems
    \item \textbf{Program:} Python framework for LLM energy/latency modeling.
    \item \textbf{Data set:} JSON validation files, YAML configurations.
    \item \textbf{Run-time environment:} Ubuntu 22.04, NVIDIA Jetpack 6.2, CUDA 12.8, Docker, VLLM, PyYAML, NumPy.
    \item \textbf{Hardware:} NVIDIA Jetson Orin AGX 64GB, H100, RTX A6000.
    \item \textbf{Disk space required:} $\sim$64GB
    \item \textbf{How much time is needed to prepare workflow?} $\sim$30 mins.
    \item \textbf{How much time is needed to complete experiments?} $\sim$24 hours.
    \item \textbf{Publicly available?:} Yes.
    \item \textbf{Code licenses:} BSD-3-Clause license.
    \item \textbf{Workflow automation:} GNU Make + Bash scripts, Jupyter Notebook.
 \item \textbf{Archived:} \url{https://doi.org/10.5281/zenodo.17168238}
\end{itemize}

\subsection{Installation}

To set up the artifact, clone the repository and use the provided Make commands to setup for tegra or a server system.
\begin{lstlisting}[language=bash]
git clone \
https://github.com/edge-inference \
edgereasoning.git
cd edgereasoning
make venv
source .venv/bin/activate
make setup
\end{lstlisting}

\subsection{Experiment workflow}
\label{app:experiment}
The evaluation code is under \path{edgereasoning/eval/}
and splits between server and Tegra hosts.

\textbf{Tegra:}
On Tegra systems one can call the following benchmarks to produce prefill and decode data used in figures 1-5

\begin{lstlisting}[language=bash]
#enter container environment
cd eval/tegra && ./open.sh 1
./launch.sh prefill
./launch.sh decode
\end{lstlisting}

The framework consists of a benchmarking suite and analytical models for latency, power, and energy.

\textbf{Server:}
Server systems can run MMLU-Redux \cite{gema2024mmlu} benchmarks faster to produce accuracy results presented in figures 6-8:
\begin{lstlisting}[language=bash]
#in edgereasoning directory
make server-mmlu       # MMLU-Redux evaluation
make planner          # Planner benchmarks
\end{lstlisting}

\textbf{Configurations}: Each evaluation has configuration files such as \path{eval/tegra/mmlu/configs/} that define test runs such as 

\begin{lstlisting}[language=bash]
decode.yaml
prefill.yaml
base.yaml
scale.yaml
budget.yaml
noreasoning.yaml
\end{lstlisting}

Such configurations can be edited to produce desired test configuration of token budget, prompt style and more. 

\subsection{Evaluation and expected results}

\textbf{Post-processing:} After running the benchmarks, process the raw logs with the \texttt{token2metrics} module located at \path{edgereasoning/third_party/token2metrics}. This step aggregates per-token latency and power measurements. The following steps will produce figures 1-5 in \path{edgereasoning/outputs/}

\begin{lstlisting}[language=bash]
#inside edgereasoning/
python postprocess.py --sub-config prefill
python postprocess.py --sub-config decode
\end{lstlisting}

\textbf{Plotting:} Figures 1-5 can be produced using token2metrics by running the following

\begin{lstlisting}[language=bash]
cd third_party/token2metrics/prefillenergy/
./run.sh
cd third_party/token2metrics/decodeenergy/
./run.sh
\end{lstlisting}

\textbf{Analytical Models:} Fitting coefficients are produced along the figures files.
These coefficients can be used to update \path{edgereasoning/models/analytic.yaml}
Execute the following commands to test the analytical latency and energy prediction models for the Tegra device.
\begin{lstlisting}[language=bash]
python latency_model.py -i 128 -o 128
python energy_model.py -i 128 -o 128
python energy_model.py --help

\end{lstlisting}

A successful run creates a summary table of latency, power, and energy metrics. Passing \texttt{--verbose} (\texttt{-v}) prints the predicted–empirical differences using the raw validation data under \path{edgereasoning/validation}.

\textbf{Notebook}: For convenience, \path{edgereasoning/notebook.ipynb} mirrors the full workflow and can be executed end-to-end to reproduce the evaluation and analytical estimates.

\subsection{Methodology}

Submission, reviewing and badging methodology:

\begin{itemize}
  \item \url{https://www.acm.org/publications/policies/artifact-review-and-badging-current}
  \item \url{https://cTuning.org/ae}
\end{itemize}
 %
% File: 7_2appendix.tex
\clearpage
\onecolumn
\section{Evaluations Results Appendix}

Table ~\ref{tab:decode_fit_params_base} and ~\ref{tab:mmlu_budgeted} details model performance (accuracy, latency, and cost) on the 3000-question MMLU-Redux benchmark~\cite{gema2024mmlu} shown in Figs.\ref{fig:budget_overview},\ref{fig:latency_overview}, and~\ref{fig:cost_overview}.
\begin{table*}[ht]
  \centering
  \caption{MMLU-Redux — Base, Quantized (LLMC-AWQ-W4), and Direct (3k samples/row).}
  \label{tab:mmlu_base_quant_direct}
  \sisetup{
    table-number-alignment = center,
    detect-weight, detect-inline-weight,
    group-separator = {,}, group-minimum-digits = 4,
    round-mode = places
  }
  \setlength{\tabcolsep}{3pt}
  \renewcommand{\arraystretch}{0.95}
  \footnotesize
  \makebox[\textwidth][c]{%
  \begin{tabularx}{0.90\textwidth}{@{}l >{\raggedright\arraybackslash}X l
      S[table-format=2.1, round-precision=1]   % Acc %
      S[table-format=4.1, round-precision=1]   % Avg toks
      S[table-format=3.2, round-precision=2]   % Avg latency (s)
      S[table-format=1.3, round-precision=3]@{}} % Cost ($/1M toks)
    \toprule
    \textbf{Family} & \textbf{Model} & \textbf{Config} &
    {\textbf{Acc. (\%)}} & {\textbf{Avg toks/question}} & {\textbf{Avg Latency (s)}} & {\textbf{Cost (\$/1M toks)}} \\
    \midrule
    % ---- Base ----
    Base      & DSR1-Qwen-1.5B      & Distilled & 38.3 &  740.2 &  18.92 & 0.024 \\
    Base      & DSR1-Llama-8B       & Distilled & 61.7 &  811.1 &  87.16 & 0.111 \\
    Base      & DSR1-Qwen-14B       & Distilled & 80.6 & 1317.8 & 259.02 & 0.215 \\
    Base      & L1-Max              & Distilled & 43.8 &  312.6 &   7.50 & 0.013 \\
    \midrule
    % ---- Quantized ----
    Quantized & DSR1-Qwen-1.5B     & LLMC-AWQ-W4 & 37.9 &  698.5 &  9.9294 & 0.014760909 \\
    Quantized & DSR1-Llama-8B      & LLMC-AWQ-W4 & 57.9 &  549.1 & 14.6862 & 0.05282899  \\
    Quantized & DSR1-Qwen-14B       & LLMC-AWQ-W4 & 80.1 & 1235.8 & \multicolumn{1}{c}{—} & \multicolumn{1}{c}{—} \\
    \midrule
    % ---- Direct ----
    Direct    & Qwen2.5-7B-it       & Direct      & 60.9 &   40.2 &   4.26 & 0.019 \\
    Direct    & Gemma-7B-it         & Direct      & 33.9 &   44.7 &   4.71 & 0.020 \\
    Direct    & Llama3.1-8B-it      & Direct      & 58.3 &   63.5 &   6.60 & 0.027 \\
    \bottomrule
  \end{tabularx}}
\end{table*}

\begin{table*}[ht]
  \centering
  \caption{MMLU-Redux — Budgeted decoding (Hard/Soft/NR). \emph{T}=hard limit; \emph{NC}=soft limit (natural completion).}
  \label{tab:mmlu_budgeted}
  \sisetup{
    table-number-alignment = center,
    detect-weight, detect-inline-weight,
    group-separator = {,}, group-minimum-digits = 4,
    round-mode = places
  }
  \setlength{\tabcolsep}{3pt}
  \renewcommand{\arraystretch}{0.95}
  \footnotesize
    \begin{tabularx}{0.90\textwidth}{@{}X X X
         S[table-format=2.1, round-precision=1]  % Accuracy (%)
         S[table-format=4.1, round-precision=1]  % Avg tokens
         S[table-format=3.3, round-precision=3]  % Avg latency (s)
         S[table-format=1.3, round-precision=3]@{}} % Cost ($/1M toks)
         \toprule
         \textbf{Model} & \textbf{BudgetType} & \textbf{ConfigLabel} &
         {\textbf{Acc. (\%)}} & {\textbf{Avg toks/question}} & {\textbf{Avg Latency (s)}} & {\textbf{Cost (\$/1M toks)}} \\
         \midrule
         % -- DSR1-Llama-8B --
         DSR1-Llama-8B  & Soft & 128 (NC)  & 60.4 &  437.0  &  46.939 & 0.096 \\
         DSR1-Llama-8B  & Soft & 256 (NC)  & 64.3 &  933.0  &  97.908 & 0.109 \\
         DSR1-Llama-8B  & NR   & NR        & 51.0 &  182.94 &  18.661 & 0.061 \\
         DSR1-Llama-8B  & Hard & 128T      & 37.9 &   76.3  &   7.888 & 0.031 \\
         DSR1-Llama-8B  & Hard & 256T      & 41.2 &  143.6  &  14.661 & 0.048 \\
         \midrule
         % -- DSR1-Qwen-1.5B --
         DSR1-Qwen-1.5B & Soft & 128 (NC)  & 35.5 & 1474.0  &  38.001 & 0.028 \\
         DSR1-Qwen-1.5B & Soft & 256 (NC)  & 39.4 &  734.8  &  18.175 & 0.021 \\
         DSR1-Qwen-1.5B & NR   & NR        & 41.0 &  234.86 &   5.644 & 0.012 \\
         DSR1-Qwen-1.5B & Hard & 128T      & 15.9 &   91.5  &   2.221 & 0.005 \\
         DSR1-Qwen-1.5B & Hard & 256T      & 23.2 &  144.1  &   3.468 & 0.007 \\
         \midrule
         % -- DSR1-Qwen-14B --
         DSR1-Qwen-14B  & Soft & 128 (NC)  & 76.9 &  599.0  & 118.091 & 0.189 \\
         DSR1-Qwen-14B  & Soft & 256 (NC)  & 77.2 &  374.2  &  70.917 & 0.152 \\
         DSR1-Qwen-14B  & NR   & NR        & 69.0 &  180.72 &  34.201 & 0.115 \\
         DSR1-Qwen-14B  & Hard & 128T      & 46.1 &   78.2  &  15.013 & 0.064 \\
         DSR1-Qwen-14B  & Hard & 256T      & 58.6 &  112.9  &  21.485 & 0.082 \\
         \midrule
         % -- L1-Max --
         L1-Max         & Soft & 128 (NC)  & 17.8 &   54.3  &   1.353 & 0.004 \\
         L1-Max         & Soft & 256 (NC)  & 17.1 &   62.3  &   1.552 & 0.005 \\
         L1-Max         & Hard & 128T      & 16.2 &   40.7  &   1.019 & 0.003 \\
         L1-Max         & Hard & 256T      & 18.3 &   48.9  &   1.213 & 0.003 \\
         \bottomrule
       \end{tabularx}
\end{table*}

%table for the CAIS original mmlu performance
\clearpage
Table ~\ref{tab:mmlu_performance} lists the additional model performance evaluation (accuracy, latency, and cost) on the MMLU benchmark~\cite{hendrycks2021mmlu} with 15k questions.

\begin{table*}[ht]
  \centering
  \caption{MMLU~\cite{hendrycks2021mmlu} accuracy (15k questions) for base, quantized, and budgeted DSR1 models.}
  \label{tab:mmlu_performance}
  \sisetup{
    table-number-alignment = center,
    detect-weight = true,
    round-mode = places
  }
  \setlength{\tabcolsep}{3pt}
  \renewcommand{\arraystretch}{0.95}
  \footnotesize
  \makebox[\textwidth][c]{%
  \begin{tabularx}{0.90\textwidth}{@{}l
      >{\raggedright\arraybackslash}X
      S[table-format=2.2, round-precision=2]   % Accuracy (%)
      S[table-format=4.1, round-precision=1]@{}} % Avg toks/q
    \toprule
    \textbf{Model} & \textbf{Configuration} & {\textbf{Accuracy (\%)}} & {\textbf{Avg toks/q}} \\
    \midrule

    % ===== 1.5B =====
    \multicolumn{4}{@{}l}{\textbf{DSR1-Qwen-1.5B}}\\
     & Base                   & 41.67 & 1141.56 \\
     & \quad Budget 128T      & 24.60 &   88.67 \\
     & \quad Budget 256T      & 29.60 &  113.68 \\
    \cmidrule(l){2-4}
     & LLMC-AWQ-W4            & 37.73 &  984.37 \\
     & \quad Budget 128T      & 24.60 &   86.87 \\
     & \quad Budget 256T      & 29.10 &  120.39 \\
    \midrule

    % ===== 8B =====
    \multicolumn{4}{@{}l}{\textbf{DSR1-Llama-8B}}\\
     & Base                   & 60.38 &  345.55 \\
     & \quad Budget 128T      & 31.03 &  101.52 \\
     & \quad Budget 256T      & 41.80 &  169.34 \\
    \cmidrule(l){2-4}
     & LLMC-AWQ-W4            & 60.44 &  455.42 \\
     & \quad Budget 128T      & 32.10 &   97.65 \\
     & \quad Budget 256T      & 43.50 &  157.05 \\
    \midrule

    % ===== 14B =====
    \multicolumn{4}{@{}l}{\textbf{DSR1-Qwen-14B}}\\
     & Base                   & 86.59 & 1145.38 \\
     & \quad Budget 128T      & 28.30 &  193.35 \\
     & \quad Budget 256T      & 37.70 &  185.69 \\
    \cmidrule(l){2-4}
     & LLMC-AWQ-W4            & 86.69 & 1148.41 \\
     & \quad Budget 128T      & 27.10 &  109.57 \\
     & \quad Budget 256T      & 37.10 &  162.00 \\
    \bottomrule
  \end{tabularx}}%
\end{table*}

Tables~\ref{tab:natplan_baseline}, ~\ref{tab:natplan_budget}, and ~\ref{tab:natplan_direct} show the model performance evaluation (accuracy and  latency) on the
Natural-Plan benchmark~\cite{zheng2024natural}.

% ---- Baseline: reasoning models ----
\begin{table}[H]
  \centering
  % \begin{subtable}{0.4\linewidth}
  \begin{minipage}{0.45\linewidth}
  \centering
  
  \caption{Baseline (reasoning models)}
  \label{tab:natplan_baseline}
  \sisetup{table-number-alignment=center, detect-weight=true, detect-inline-weight=math}
  \setlength{\tabcolsep}{3pt}
\resizebox{0.8\columnwidth}{!} {
  
  \begin{tabular}{@{}l l S[table-format=2.2] S[table-format=4.0] S[table-format=2.2]@{}}
    \toprule
    \textbf{Task} & \textbf{Model} & {\textbf{Acc. (\%)}} & {\textbf{Avg out toks/Q}} & {\textbf{Lat. (s)}} \\
    \midrule
    % 1.5B
    calendar & 1.5B &  0.60 & 2792 &  8.90 \\
    meeting  & 1.5B &  1.00 & 3880 & 19.90 \\
    trip     & 1.5B &  1.25 & 2490 &  7.88 \\
    \midrule
    % 8B
    calendar & 8B   &  9.00 & 2798 & 21.10 \\
    meeting  & 8B   & 10.00 & 2866 & 24.50 \\
    trip     & 8B   &  7.88 & 2251 & 17.10 \\
    \midrule
    % 14B
    calendar & 14B  & 11.70 & 2297 & 30.00 \\
    meeting  & 14B  & 19.30 & 1494 & 22.10 \\
    trip     & 14B  & 13.88 & 2340 & 30.40 \\
    \bottomrule
  \end{tabular}}
  \end{minipage}\hfill
% \end{table}
% ---- Direct Models (Qwen2.5) ----
% \begin{table}[H]
  \begin{minipage}{0.45\linewidth}  

  \centering
  \caption{Budgeting (NR + Hard limit at 512 tokens)}
  \label{tab:natplan_budget}
  \sisetup{table-number-alignment=center, detect-weight=true, detect-inline-weight=math}
  \setlength{\tabcolsep}{3pt}
  \resizebox{0.8\columnwidth}{!} {
  
  \begin{tabular}{@{}l l S[table-format=2.2] S[table-format=3.0] S[table-format=1.3]@{}}
    \toprule
    \textbf{Task} & \textbf{Model} & {\textbf{Acc. (\%)}} & {\textbf{Avg out toks/Q}} & {\textbf{Lat. (s)}} \\
    \midrule
    % 1.5B
    calendar & 1.5B &  2.00 & 511 & 2.840 \\
    meeting  & 1.5B &  1.90 & 425 & 1.350 \\
    trip     & 1.5B &  0.00 & 507 & 1.420 \\
    \midrule
    % 8B
    calendar & 8B   &  8.10 &  67 & 0.552 \\
    meeting  & 8B   & 11.90 & 284 & 2.510 \\
    trip     & 8B   &  3.90 & 398 & 3.094 \\
    \midrule
    % 14B
    calendar & 14B  & 12.60 &  40 & 0.615 \\
    meeting  & 14B  & 19.00 & 341 & 5.223 \\
    trip     & 14B  & 10.90 & 380 & 4.984 \\
    \bottomrule
  \end{tabular}}
  \end{minipage}  
\end{table}

% ---- Budgeting: NR + 512T ----
\begin{table}[H]
  \centering
  \caption{Direct models (Qwen2.5)}
  \label{tab:natplan_direct}
  \sisetup{table-number-alignment=center, detect-weight=true, detect-inline-weight=math}
  \setlength{\tabcolsep}{3pt}

  \begin{tabular}{@{}l l S[table-format=2.2] S[table-format=3.0] S[table-format=1.3]@{}}
    \toprule
    \textbf{Task} & \textbf{Model sz.} & {\textbf{Acc. (\%)}} & {\textbf{Avg out toks/Q}} & {\textbf{Lat. (s)}} \\
    \midrule
    % 1.5B
    calendar & 1.5B &  5.30 &  22 & 0.087 \\
    meeting  & 1.5B &  9.40 & 271 & 1.369 \\
    trip     & 1.5B &  2.50 & 242 & 0.804 \\
    \midrule
    % 14B
    calendar & 14B  & 31.90 &  28 & 0.464 \\
    meeting  & 14B  & 27.20 & 283 & 4.408 \\
    trip     & 14B  &  6.44 & 259 & 3.440 \\
    \bottomrule
  \end{tabular}
\end{table}

\clearpage
% File: 7_2appendix.tex
\clearpage
\twocolumn
% \lstdefinestyle{codebox}{%
%   frame=single,
%   rulecolor=\color{black!50},
%   framerule=0.6pt,
%   backgroundcolor=\color{black!3},
%   basicstyle=\ttfamily\small,
%   numbers=none,
%   xleftmargin=1em,
%   xrightmargin=1em,
%   framexleftmargin=1em,
%   breaklines=false,      
%   breakatwhitespace=false,
%   columns=fullflexible,
%   keepspaces=true,
%   showstringspaces=false,
%   tabsize=2
% }
% \lstset{style=codebox}

% \section{Additional Evaluations Appendix: EdgeReasoning}
\section{Edge CPU Evaluation}
% \subsection{Additional Evaluations}
This section presents characterization results for a 12-core Arm Cortex-A78AE CPU, evaluated as an alternative inference platform.
% \FloatBarrier

% \subsection{CPU and GPU Results Comparison}
% \FloatBarrier
\begin{table}[H]
  \centering
  \caption{Prefill Latency: CPU vs. GPU}
  \label{tab:latency_by_inputlen}
  \sisetup{table-number-alignment = center, detect-weight=true, detect-inline-weight=math}
  \setlength{\tabcolsep}{4pt}
  \begin{tabular}{
    @{}r 
    S[table-format=3.2] S[table-format=1.3]
    S[table-format=3.1] S[table-format=1.3]
    S[table-format=3.2] S[table-format=1.3]@{}
  }
    \toprule
    \multicolumn{1}{r}{\scriptsize\textbf{Len}} &
      \multicolumn{2}{c}{\textbf{1.5B}} &
      \multicolumn{2}{c}{\textbf{8B}} &
      \multicolumn{2}{c}{\textbf{14B}} \\
    \cmidrule(lr){2-3}\cmidrule(lr){4-5}\cmidrule(lr){6-7}
    & {\textbf{CPU (s)}} & {\textbf{GPU (s)}}
    & {\textbf{CPU (s)}} & {\textbf{GPU (s)}}
    & {\textbf{CPU (s)}} & {\textbf{GPU (s)}} \\
    \midrule
    128  &  8.44 & 0.051 &  46.5 & 0.148 &  79.29 & 0.270 \\
    256  & 17.0  & 0.054 &  89.7 & 0.223 & 167.0  & 0.421 \\
    512  & 37.1  & 0.095 & 157   & 0.554 & 344.2  & 0.764 \\
    1024 & 75.6  & 0.158 & 384   & 0.801 & 734.2  & 1.521 \\
    \bottomrule
  \end{tabular}
\end{table}

\begin{table}[H]
  \centering
  \caption{Decode Latency: CPU vs. GPU}
  \label{tab:decode_latency_512in}
  \sisetup{table-number-alignment=center, detect-weight=true, detect-inline-weight=math}
  \setlength{\tabcolsep}{4pt} 
  \begin{tabular}{
    @{}r 
    S[table-format=3.1] S[table-format=3.1]
    S[table-format=3.1] S[table-format=3.1]@{}
  }
    \toprule
    \multicolumn{1}{r}{\scriptsize\textbf{Output Length}} &
      \multicolumn{2}{c}{\textbf{8B}} &
      \multicolumn{2}{c}{\textbf{14B}} \\
    \cmidrule(lr){2-3}\cmidrule(lr){4-5}
     & {\textbf{CPU (s)}} & {\textbf{GPU (s)}} & {\textbf{CPU (s)}} & {\textbf{GPU (s)}} \\
    \midrule
    64   & 259.9 &  52.1 & 461.7 &  95.3 \\
    128  &  63.8 &  12.9 & 113.5 &  23.7 \\
    256  & 128.8 &  26.1 & 228.8 &  47.5 \\
    1024 & 521.5 & 104.5 & 926.5 & 190.5 \\
    \bottomrule
  \end{tabular}
\end{table}

\newpage
\section{Quantized Models Evaluation}

Tables~\ref{tab:prefill_base_vs_awq} and~\ref{tab:filtered512} present a performance comparison between the base FP16 models and their W4A16-quantized counterparts.

%quantization

% Prefill: Base vs Quantized (AWQ W4)
\begin{table}[htbp]
  \centering
  \caption{Prefill Performance: Base vs Quantized. Averaged across input length sweep range [128, 4096] }
  \label{tab:prefill_base_vs_awq}
  \sisetup{table-number-alignment=center, round-mode=places}
  \setlength{\tabcolsep}{2.5pt}
  \renewcommand{\arraystretch}{0.95}
  \scriptsize
  \begin{tabularx}{\columnwidth}{
    @{}>{\raggedright\arraybackslash}X
    S[table-format=4.2, round-precision=2]   % Time (s)
    S[table-format=2.1, round-precision=1]   % Tok/s
    S[table-format=2.1, round-precision=1]@{}} % Power (W)
    \toprule
    \textbf{Model} & {\textbf{Time (s)}} & {\textbf{Tok/s}} & {\textbf{Power (W)}} \\
    \midrule
    \multicolumn{4}{@{}l}{\textbf{Base}}\\
    DSR1-Qwen-1.5B  & 0.33 & 5.3 &  5.6 \\
    DSR1-Llama-8B   & 2.60 & 1.2 & 17.0 \\
    DSR1-Qwen-14B   & 3.63 & 0.7 & 23.5 \\
    \midrule
    \multicolumn{4}{@{}l}{\textbf{Quantized (AWQ W4)}}\\
    DSR1-1.5B--AWQ-W4   & 0.15 & 9.8 &  4.8 \\
    DSR1-8B--AWQ-W4     & 0.55 & 5.1 & 13.6 \\
    DSR1-14B--AWQ-W4    & 2.21 & 1.8 & 20.5 \\
    \bottomrule
  \end{tabularx}
\end{table}

% Model Summary @ 512 input tokens (Time=2dp, Tok/s=1dp, Power=1dp)
\begin{table}[htbp]
  \centering
  \caption{Decode Performance: Base vs Quantized Decode. Input length 512: and output length sweep range: [128, 2048]}
  \label{tab:filtered512}
  \sisetup{table-number-alignment=center, round-mode=places}
  \setlength{\tabcolsep}{2.5pt}
  \renewcommand{\arraystretch}{0.95}
  \scriptsize
  \begin{tabularx}{\columnwidth}{
    @{}>{\raggedright\arraybackslash}X
    S[table-format=3.2, round-precision=2] % Time (s): 2 dp
    S[table-format=3.1, round-precision=1] % Tok/s: 1 dp
    S[table-format=2.1, round-precision=1]@{}} % Power (W): 1 dp
    \toprule
    \textbf{Model} & {\textbf{Time (s)}} & {\textbf{Tok/s}} & {\textbf{Power (W)}} \\
    \midrule
    \multicolumn{4}{@{}l}{\textbf{Base (distilled)}}\\
    DeepSeek-R1-Distill-Qwen-1.5B  & 20.858 & 38.2011 & 19.6288 \\
    DeepSeek-R1-Distill-Llama-8B   & 86.418 &  9.0482 & 24.3542 \\
    DeepSeek-R1-Distill-Qwen-14B   &158.178 &  4.9927 & 26.5254 \\
    \midrule
    \multicolumn{4}{@{}l}{\textbf{Quantized (AWQ W4)}}\\
    DSR1-1.5B-llmc-awq-w4          & 10.643 & 73.5559 & 16.2274 \\
    DSR1-8B-llmc-awq-w4            & 29.937 & 25.8678 & 25.4413 \\
    DSR1-14B-llmc-awq-w4           & 51.055 & 15.1025 & 28.4830 \\
    \bottomrule
  \end{tabularx}
\end{table}
\clearpage
% File: 7_2appendix.tex

\lstdefinestyle{codebox}{%
  frame=single,
  rulecolor=\color{black!50},
  framerule=0.6pt,
  backgroundcolor=\color{black!3},
  basicstyle=\ttfamily\small,
  numbers=none,
  xleftmargin=1em,
  xrightmargin=1em,
  framexleftmargin=1em,
  breaklines=false,      
  breakatwhitespace=false,
  columns=fullflexible,
  keepspaces=true,
  showstringspaces=false,
  tabsize=2
}
\lstset{style=codebox}
\clearpage
\onecolumn
\section{Appendix: EdgeReasoning}
\subsection{Fitted Coefficients for Energy and Power Modeling}
% Prefill power/energy fit coefficients (DeepSeek R1 distilled)
\begin{table*}[h]
\centering
\caption{Fitted parameters for \emph{prefill} power and energy models (DeepSeek R1 distilled). \(I\): input length in tokens.}
\label{tab:prefill_fit_params_base}
\resizebox{\textwidth}{!}{%
\begin{tabular}{llll}
\toprule
\textbf{Model} & \textbf{Power Function} & \textbf{Energy Function} & \textbf{Key Parameters} \\
\midrule
1.5B & Constant: $P=5.636$ 
     & Exp.\ decay: $E = A e^{-\lambda I} + C$
     & $A=0.07308$, $\lambda=0.03195$, $C=0.000923$ \\
\addlinespace
8B   & Const.\ ($I\le 800$), Log ($I>800$)
     & Piecewise: Exp.\ decay ($I\le 640$), Log ($I>640$)
     & Exp: $A=0.15871$, $\lambda=0.03240$, $C=0.00553$; \quad Log: $\alpha=0.01233$, $\beta=-0.07349$ \\
\addlinespace
14B  & Const.\ ($I\le 384$), Log ($I>384$)
     & Piecewise: Exp.\ decay ($I\le 384$), Log ($I>384$)
     & Exp: $A=0.29327$, $\lambda=0.03058$, $C=0.009234$; \quad Log: $\alpha=0.01605$, $\beta=-0.07643$ \\
\bottomrule
\end{tabular}}
\end{table*}

% Decode power/energy fit coefficients (DeepSeek R1 distilled)
\begin{table*}[h]
\centering
\caption{Fitted parameters for \emph{decode} power and energy models (distilled models). \(O\) output length in tokens.}
\label{tab:decode_fit_params_base}
\resizebox{\textwidth}{!}{%
\begin{tabular}{llll}
\toprule
\textbf{Model} & \textbf{Power Function} & \textbf{Energy Function} & \textbf{Key Parameters} \\
\midrule
1.5B &
Log: $P=\alpha \ln O + \beta$ &
Log: $E=\alpha \ln O + \beta$ &
Power: $\alpha=0.756538$, $\beta=3.213711$;\quad Energy: $\alpha=-0.059992$, $\beta=0.091465$ \\
\addlinespace
8B &
Log: $P=\alpha \ln O + \beta$ &
Log: $E=\alpha \ln O + \beta$ &
Power: $\alpha=8.806744$, $\beta=2.701709$;\quad Energy: $\alpha=0.555184$, $\beta=0.324112$ \\
\addlinespace
14B &
Log: $P=\alpha \ln O + \beta$ &
Log: $E=\alpha \ln O + \beta$ &
Power: $\alpha=16.886830$, $\beta=1.619387$;\quad Energy: $\alpha=1.764896$, $\beta=0.515518$ \\
\bottomrule
\end{tabular}}
\end{table*}

% Prefill:  Quantized models
\begin{table*}[h]
\centering
\caption{Fitted parameters for prefill power and energy models for quantized models. \(I\): input length in tokens.}
\label{tab:prefill_fit_params}
\resizebox{\textwidth}{!}{%
\begin{tabular}{llll}
\toprule
Model & Power Function & Energy Function & Key Parameters \\
\midrule
1.5B & Constant: $P=4.83$ & Exp. decay: $E=0.093 e^{-0.109N}+0.0011$ 
& $A=0.093$, $\lambda=0.109$, $C=0.0011$ \\
\addlinespace
14B & Const. ($I\le384$), Log ($I>384$) 
& Piecewise: Exp. decay ($I\le640$), Log ($I>640$) 
& Exp: $A=0.160$, $\lambda=0.129$, $C=0.008$; Log: $\alpha=0.0157$, $\beta=-0.089$ \\
\addlinespace
8B & Const. ($I\le1400$), Log ($I>1400$) 
& Piecewise: Exp. decay ($I\le1500$), Log ($I>1500$) 
& Exp: $A=0.101$, $\lambda=0.121$, $C=0.0037$; Log: $\alpha=0.0066$, $\beta=-0.040$ \\
\bottomrule
\end{tabular}}
\end{table*}

\begin{table*}[h]
\centering
\caption{Fitted parameters for \emph{decode} power and energy models (quantized W4). \(O\): output length in tokens}
\label{tab:decode_fit_params_quant_w4}
\resizebox{\textwidth}{!}{%
\begin{tabular}{llll}
\toprule
\textbf{Model} & \textbf{Power Function} & \textbf{Energy Function} & \textbf{Key Parameters} \\
\midrule
DSR1\text{-}Qwen\text{-}1.5B\text{-}W4 &
Log: $P=\alpha \ln O + \beta$ &
Log: $E=\alpha \ln O + \beta$ &
Power: $\alpha=3.0401$, $\beta=-1.6672$;\quad Energy: $\alpha=0.04338$, $\beta=-0.05468$ \\
\addlinespace
DSR1\text{-}Llama\text{-}8B\text{-}W4 &
Log: $P=\alpha \ln O + \beta$ &
Log: $E=\alpha \ln O + \beta$ &
Power: $\alpha=3.8723$, $\beta=3.0186$;\quad Energy: $\alpha=0.15962$, $\beta=-0.05413$ \\
\addlinespace
DSR1\text{-}Qwen\text{-}14B\text{-}W4 &
Log: $P=\alpha \ln O + \beta$ &
Log: $E=\alpha \ln O + \beta$ &
Power: $\alpha=3.0515$, $\beta=11.0898$;\quad Energy: $\alpha=0.24460$, $\beta=0.24737$ \\
\bottomrule
\end{tabular}}
\end{table*}

 %
% % \section*{Acknowledgment}
% % \vspace{12pt}
\bibliographystyle{plain}
\clearpage

\end{document}